\documentclass[reqno]{amsart}

\usepackage{booktabs} 
\usepackage{IEEEtrantools}
\usepackage{amssymb,latexsym,amsfonts,amsmath}
\usepackage{graphicx}
\usepackage{mathrsfs}
\usepackage{dsfont}

\usepackage{tikz}
\usetikzlibrary{calc,shapes,arrows}

\usepackage{algorithm}
\usepackage{algorithmic}

\usepackage{xspace}
\newcommand{\software}{\textsf{FAUST}$^{\mathsf 2}$\xspace}

\newtheorem{theorem}{Theorem}[section]
\newtheorem{lemma}[theorem]{Lemma}

\newtheorem{proposition}[theorem]{Proposition}

\newtheorem{definition}[theorem]{Definition}

\newtheorem{remark}[theorem]{Remark}
\newtheorem{assumption}{Assumption}
\numberwithin{equation}{section}

\newcommand{\R}{{\mathbb{R}}}

\newcommand{\N}{{\mathbb{N}}}

\newcommand{\PP}{\mathds{P}}

\begin{document}

\begin{abstract}
In this paper, we provide a compositional approach for constructing finite abstractions (a.k.a. finite Markov decision processes (MDPs)) of interconnected discrete-time stochastic switched systems. The proposed framework is based on a notion of \emph{stochastic simulation functions}, using which one can employ an abstract system as a substitution of the original one in the controller design process with guaranteed error bounds on their output trajectories. To this end, we first provide probabilistic closeness guarantees between the interconnection of stochastic switched subsystems and that of their finite abstractions via stochastic simulation functions. We then leverage sufficient small-gain type conditions to show compositionality results of this work. Afterwards, we show that under standard assumptions ensuring incremental input-to-state stability of switched systems (i.e., existence of \emph{common} incremental Lyapunov functions, or \emph{multiple} incremental Lyapunov functions with \emph{dwell-time}), one can construct finite MDPs for the general setting of nonlinear stochastic switched systems. We also propose an approach to construct finite MDPs together with their corresponding stochastic simulation functions for a particular class of nonlinear stochastic switched systems. We show that for this class of systems, the aforementioned incremental stability property can be readily checked by matrix inequalities. To demonstrate the effectiveness of our proposed results, we first apply our approaches to a road traffic network in a circular cascade ring composed of $200$ cells, and construct compositionally a finite MDP of the network. We employ the constructed finite abstractions as substitutes to compositionally synthesize policies keeping the density of the traffic lower than $20$ vehicles per cell. We then apply our proposed techniques to a \emph{fully interconnected} network of $500$ nonlinear subsystems (totally $1000$ dimensions), and construct their finite MDPs with guaranteed error bounds. We compare our proposed results with those available in the literature.
\end{abstract}

\title[Compositional Abstraction-based Synthesis for Networks of Stochastic Switched Systems]{Compositional Abstraction-based Synthesis for Networks of Stochastic Switched Systems}

\author{Abolfazl Lavaei$^1$}
\author{Sadegh Soudjani$^2$}
\author{Majid Zamani$^{3,1}$}
\address{$^1$Department of Computer Science, Ludwig Maximilian University of Munich, Germany.}
\email{lavaei@lmu.de}
\address{$^2$School of Computing, Newcastle University, UK.}
\email{sadegh.soudjani@ncl.ac.uk}
\address{$^3$Department of Computer Science, University of Colorado Boulder, USA.}
\email{majid.zamani@colorado.edu}
\maketitle

\section{Introduction}
{\bf Motivations.} In recent years, switched systems as an important modeling framework describing many engineering systems have received significant attentions due to their broad presence in real-life applications. It is understood that by fast switching between even stable subsystems, one may render the overall system unstable. This issue motivated many researchers over the past few years to investigate mainly which classes of switching strategies or switching signals preserve stability~\cite{liberzon2003switching}.

In the past few years, there have been many works on the synthesis of controllers rendering switched systems stable. However, there is only a limited work on the construction of controllers for such systems with respect to complex logic properties. In fact, automated controller synthesis for complex switched systems to achieve some high-level specifications, e.g. those expressed as linear temporal logic (LTL) formulae~\cite{pnueli1977temporal}, is inherently very challenging. To tackle this complexity, one promising approach is to employ finite abstractions of the given systems as a replacement in the controller synthesis procedure. In this regard, one can first abstract the original system by a simpler one (with finite-state set), perform analysis and synthesis over the abstract model (using algorithmic techniques from computer science~\cite{baier2008principles}), and finally carry the results back over the concrete system, by providing guaranteed error bounds in this detour process.

One of the main challenges in the construction of finite abstractions for large-scale complex systems is the curse of dimensionality: the complexity  grows exponentially with the dimension of the state set. Then compositional abstraction-based techniques are essential to alleviate this complexity. In this respect, one needs to consider the large-scale switched system as an interconnected system composed of several smaller subsystems, and provide a compositional framework for the construction of finite abstractions for the given system using abstractions of smaller subsystems.

There have been several results, proposed in the past few years, on the construction of (in)finite abstractions for stochastic systems. Existing results include finite bisimilar abstractions for randomly switched stochastic systems~\cite{zamani2014approximately}, incrementally stable stochastic switched systems~\cite{zamani2015symbolic}, and stochastic control systems without discrete dynamics~\cite{zamani2014symbolic}. Infinite approximation techniques for jump-diffusion systems are also presented in~\cite{julius2009approximations}. In addition, compositional construction of infinite abstractions for jump-diffusion systems using small-gain type conditions is discussed in~\cite{zamani2016approximations}. Construction of finite abstractions for formal verification and synthesis is initially proposed in~\cite{APLS08}. Extension of such techniques to automata-based controller synthesis, and improvement of the construction algorithms in terms of scalability are proposed in~\cite{Kamgarpour2013}, and~\cite{SA13}, respectively. 

The formal abstraction-based policy synthesis is discussed in~\cite{tmka2013}, and the extension of such techniques to infinite horizon properties is discussed in~\cite{tkachev2011infinite}. Compositional construction of finite abstractions is presented in~\cite{SAM17,lavaei2018ADHS} using respectively dynamic Bayesian networks and small-gain type conditions. Compositional infinite and finite abstractions in a unified framework via approximate probabilistic relations are proposed in~\cite{lavaeiNSV2019,lavaei2019NAHS1}. Compositional construction of finite MDPs for large-scale stochastic switched systems via a dissipativity approach is presented in~\cite{lavaei2019LSS}. Compositional construction of finite abstractions for networks of not necessarily stabilizable stochastic systems via relaxed small-gain and dissipativity conditions is respectively discussed in~\cite{lavaei2019ECC,lavaei2019NAHS}. An (in)finite abstraction-based technique for synthesis of stochastic control systems is recently studied in~\cite{Amy2019}.

There have been also several results on compositional verification of stochastic models. Similarity relations over finite-state stochastic systems have been studied either via exact notions of probabilistic (bi)simulation relations~\cite{larsen1991bisimulation},~\cite{segala1995probabilistic}, or approximate versions~\cite{desharnais2008approximate},~\cite{d2012robust}. Compositional modelling and analysis for the safety verification of stochastic hybrid systems are investigated in~\cite{hahn2013compositional} in which random behaviour occurs only over the discrete components. Compositional controller synthesis for stochastic games using an assume-guarantee reasoning for the probabilistic finite automata is proposed in~\cite{basset2014compositional}. In addition, compositional probabilistic verification via an assume-guarantee framework based on multi-objective probabilistic model checking is investigated in~\cite{kwiatkowska2013compositional} for finite systems. Recently, a quantized feedback control of nonlinear Markov jump systems, and a dissipative filtering approach for a class of discrete-time switched fuzzy systems with missing measurements are proposed in~\cite{zhang2018quantized}, and~\cite{zhang2019dissipative}, respectively.

{\bf Our Contributions.} Our main contribution here is to provide for the first time a compositional methodology for the construction of finite MDPs for networks of stochastic \emph{switched} systems accepting \emph{multiple} Lyapunov functions with \emph{dwell-time}. The proposed technique leverages sufficient small-gain type conditions to establish the compositionality results which rely on relations between subsystems and their abstractions described by the existence of \emph{stochastic simulation functions}. This type of relations enables us to compute the probabilistic error between the interconnection of concrete subsystems and that of their finite abstractions. In this respect, 
we first leverage sufficient small-gain type conditions for the compositional quantification of the probabilistic distance between the interconnection of stochastic switched subsystems and that of their finite abstractions. 
We then show that under standard assumptions ensuring incremental input-to-state stability of a switched system (i.e., existence of a \emph{common} incremental Lyapunov function, or \emph{multiple} incremental Lyapunov functions with \emph{dwell-time}), one can construct finite MDPs of nonlinear stochastic switched systems.

We also propose an approach to construct finite MDPs together with their corresponding stochastic simulation functions for a particular class of \emph{nonlinear} stochastic switched systems. We show that for this class of nonlinear switched systems, the aforementioned incremental input-to-state stability property can be readily checked by matrix inequalities. To demonstrate the effectiveness of our proposed results, we first apply our approaches to a road traffic network in a circular cascade ring composed of $200$ cells, each of which has the length of $500$ meters with $1$ entry and $1$ way out, and construct compositionally a \emph{finite} MDP of the network. We employ the constructed finite abstractions as substitutes to compositionally synthesize policies keeping the density of traffic lower than $20$ vehicles per cell. Finally, we show the applicability of our results to switched systems accepting \emph{multiple} Lyapunov functions with \emph{dwell-time}. We apply our proposed techniques to a \emph{fully interconnected} network of $500$ \emph{nonlinear} subsystems (totally $1000$ dimensions) and construct their \emph{finite} MDPs with guaranteed error bounds. We compare our results with the compositional techniques proposed in~\cite{SAM17} and~\cite{lavaei2017HSCC}.

{\bf Recent Works.} Compositional construction of infinite abstractions (reduced-order models) for networks of stochastic control systems is proposed in~\cite{lavaei2017compositional} and~\cite{lavaei2018CDCJ} using small-gain type conditions and dissipativity-type properties of subsystems and their abstractions, respectively. Compositional construction of finite abstractions is presented in~\cite{lavaei2017HSCC} and~\cite{lavaei2018ADHSJJ} using respectively dissipativity-type reasoning and small-gain conditions, both for discrete-time stochastic control systems. In comparison with the current work, the proposed results  in~\cite{lavaei2017HSCC},~\cite{lavaei2017compositional},~\cite{lavaei2018CDCJ},~\cite{lavaei2018ADHSJJ}  are about the compositional construction of (in)finite abstractions for stochastic \emph{control} systems, while here for the first time we enlarge the class of systems to \emph{switched} ones.
If switched systems accept common Lyapunov functions, our proposed results here recover the ones presented in the previous works by considering switching signals as discrete inputs. In this respect, we make comparisons between our results with the ones proposed in~\cite{SAM17} and~\cite{lavaei2017HSCC} by providing adequate numerical implementations in the first case study. We show that our proposed results here which are based on $\max$ small-gain conditions significantly outperform the results provided in~\cite{SAM17} and~\cite{lavaei2017HSCC} which are respectively based on dynamic Bayesian network (DBN) and dissipativity-type conditions. This outperformance is due to the fact that the approximation error in~\cite{SAM17} and~\cite{lavaei2017HSCC} increases as the number of subsystems grows. Whereas, our error provided in~\eqref{Eq_25} does not change since the overall approximation error is completely independent of the size of the network, and is computed only based on the maximum error of subsystems instead of being a linear combination of them which is the case in~\cite{SAM17} and~\cite{lavaei2017HSCC}.

\section{Discrete-Time Stochastic switched Systems}

\subsection{Preliminaries}
We consider a probability space $(\Omega,\mathcal F_{\Omega},\mathds{P}_{\Omega})$,
where $\Omega$ is the sample space,
$\mathcal F_{\Omega}$ is a sigma-algebra on $\Omega$ comprising subsets of $\Omega$ as events,
and $\mathds{P}_{\Omega}$ is a probability measure that assigns probabilities to events.
We assume that random variables introduced in this article are measurable functions of the form $X:(\Omega,\mathcal F_{\Omega})\rightarrow (S_X,\mathcal F_X)$.
Any random variable $X$ induces a probability measure on  its space $(S_X,\mathcal F_X)$ as $Prob\{A\} = \mathds{P}_{\Omega}\{X^{-1}(A)\}$ for any $A\in \mathcal F_X$.
We often directly discuss the probability measure on $(S_X,\mathcal F_X)$ without explicitly mentioning the underlying probability space and the function $X$ itself.

A topological space $S$ is called a Borel space if it is homeomorphic to a Borel subset of a Polish space (i.e., a separable and completely metrizable space).
Examples of a Borel space are the Euclidean spaces $\mathbb R^n$, its Borel subsets endowed with a subspace topology, as well as hybrid spaces.
Any Borel space $S$ is assumed to be endowed with a Borel sigma-algebra, which is
denoted by $\mathcal B(S)$. We say that a map $f : S\rightarrow Y$ is measurable whenever it is Borel measurable.

\subsection{Notation}

The following notation is used throughout the paper. The sets of nonnegative and positive integers are denoted by $\mathbb N := \{0,1,2,\ldots\}$ and $\mathbb N_{\ge 1} := \{1,2,3,\ldots\}$, respectively. Moreover,
the symbols $\mathbb R$, $\mathbb R_{>0}$, and $\mathbb R_{\ge 0}$ denote, respectively, the sets of real, positive and nonnegative real numbers. For any set $X$ we denote by $2^X$ the power set of $X$ that is the set of all subsets of $X$.
Given $N$ vectors $x_i \in \mathbb R^{n_i}$, $n_i\in \mathbb N_{\ge 1}$, and $i\in\{1,\ldots,N\}$, we use $x = [x_1;\ldots;x_N]$ to denote the corresponding vector of dimension $\sum_i n_i$.
We denote by $\Vert\cdot\Vert$ and $\Vert\cdot\Vert_2$ the infinity and Euclidean norm, respectively. Symbols $\mathds{I}_n$, $\mathbf{0}_n$, and $\mathds{1}_n$ denote the identity matrix in $\mathbb R^{n\times{n}}$ and column vectors in $\mathbb R^{n\times{1}}$ with all elements equal to zero and one, respectively. The identity
function and composition of functions are denoted by $\mathcal{I}_d$ and symbol $\circ$, respectively.
Given a symmetric matrix $M$, the minimum and maximum eigenvalues of $M$ are respectively denoted by $\lambda_{\min}(M)$ and $\lambda_{\max}(M)$. We also denote by $\mathsf{diag}(a_1,\ldots,a_N)$ a diagonal matrix in $\mathbb R^{N\times{N}}$ with diagonal matrix entries $a_1,\ldots,a_N$ starting from the upper left corner. Given functions $f_i:X_i\rightarrow Y_i$,
for any $i\in\{1,\ldots,N\}$, their Cartesian product $\prod_{i=1}^{N}f_i:\prod_{i=1}^{N}X_i\rightarrow\prod_{i=1}^{N}Y_i$ is defined as $(\prod_{i=1}^{N}f_i)(x_1,\ldots,x_N)=[f_1(x_1);\ldots;f_N(x_N)]$.
For any set $\mathcal A$ we denote by $\mathcal A^{\mathbb N}$ the Cartesian product of a countable number of copies of $\mathcal A$, i.e., $\mathcal A^{\mathbb N} = \prod_{k=0}^{\infty} \mathcal A$. 
A function $\gamma:\mathbb\mathbb \mathbb R_{\ge 0}\rightarrow\mathbb\mathbb \mathbb R_{\ge 0}$, is said to be a class $\mathcal{K}$ function if it is continuous, strictly increasing, and $\gamma(0)=0$. A class $\mathcal{K}$ function $\gamma(\cdot)$ is said to be a class $\mathcal{K}_{\infty}$ if
$\gamma(r) \rightarrow \infty$ as $r\rightarrow\infty$.

\subsection{Discrete-Time Stochastic Switched Systems}
We consider stochastic switched systems in discrete-time defined formally as follows.
\begin{definition}
	A discrete-time stochastic switched system (dt-SS) is characterized by the tuple
	\begin{equation}
	\label{eq:dt-SS}
	\Sigma=(X,P,\mathcal{P},W,\varsigma, F, Y, h),
	\end{equation}
	where: 
	\begin{itemize}
		\item $X\subseteq \mathbb R^n$ is a Borel space as the state space of the system. We denote by $(X, \mathcal B (X))$ the measurable space with $\mathcal B (X)$  being  the Borel sigma-algebra on the state space;
		\item $P = \{1,\dots, m \}$  is the finite set of modes;
		\item $\mathcal{P}$ is a subset of $\mathcal{S}(\mathbb N,P)$ which denotes the set of functions from $\mathbb N$ to $P$;
		\item $W\subseteq \mathbb R^{\bar p}$ is a Borel space as the \emph{internal} input space of the system; 
		\item $\varsigma$ is a sequence of independent and identically distributed (i.i.d.) random variables from a sample space $\Omega$ to the measurable space $(\mathcal{V}_\varsigma,\mathcal F_\varsigma)$
		\begin{equation*}
		\varsigma:=\{\varsigma(k):(\Omega,\mathcal F_\Omega)\rightarrow (\mathcal{V}_\varsigma,\mathcal F_\varsigma),\,\,k\in\N\},
		\end{equation*}
		\item $F = \{f_1,\dots, f_m \}$ is a collection of vector fields indexed by $p$. For all $p\in P$, the map $f_p:X\times W\times \mathcal{V}_{\varsigma} \rightarrow X$ is a measurable function characterizing the state evolution of the system;
		\item  $Y\subseteq \mathbb R^{q}$ is a Borel space as the output space of the system;
		\item  $h:X\rightarrow Y$ is a measurable function as the output map that maps a state $x\in X$ to its output $y = h(x)$.
	\end{itemize}
	
	For a given initial state $x(0)\in X$, input sequence $w(\cdot):\mathbb N\rightarrow W$ and switching signal $\bold{p}(k):\mathbb N \rightarrow P$, evolution of the state of $\Sigma$ is described as
	\begin{equation}\label{Eq_1a}
	\Sigma:\left\{\hspace{-1.5mm}\begin{array}{l}x(k+1)=f_{\bold{p}(k)}(x(k),w(k),\varsigma(k)),\\
	y(k)=h(x(k)),\\
	\end{array}\right.
	\quad\quad k\in\mathbb N.
	\end{equation}
\end{definition}

We assume that signal $\bold{p}$ satisfies a \emph{dwell-time} condition~\cite{morse1996supervisory} as defined in the next definition.
\begin{definition}\label{dwell-time}
	Consider a switching signal $\bold{p}:\mathbb N\rightarrow P$ and define its switching time instants as
	$$\mathfrak S_{\bold{p}} := \left\{\mathfrak s_k: k\in\mathbb N_{\ge 1}\right\}\!.$$ Then, $\bold{p}:\mathbb N\rightarrow P$ has \emph{dwell-time} $k_d \in \mathbb N$~\cite{morse1996supervisory} if elements of $\mathfrak S_{\bold{p}}$ ordered as $\mathfrak s_1\le \mathfrak s_2\le \mathfrak s_3\le \dots$ satisfy $\mathfrak s_1\geq k_d$ and $\mathfrak s_{k+1} - \mathfrak s_{k} \geq k_d, \forall k\in\mathbb N_{\ge 1}$.
\end{definition}

\begin{remark}
	Note that the dwell-time  in our setting is deterministic and always respected by the controller designed using the finite MDP. More precisely, switching signals in this work are control inputs and the main goal is to synthesize them with a specific dwell-time such that outputs of original systems satisfy some high-level specifications such as safety, reachability, etc. (cf. the first case study). In existing works with the stochastic dwell-time~(e.g. \cite{battilotti2005dwell},~\cite{xiong2013stability}), switching signals are not control inputs and are randomly changing in an adversarial manner.
\end{remark}

For any $p\in P$, we use $\Sigma_p$ to refer to system~\eqref{Eq_1a} with the constant switching signal $\bold{p}(k) = p$ for all $k\in\mathbb N$. System $\Sigma$ is called finite if $ X, W$ are finite sets and infinite otherwise.

We assume that the output map $h$ satisfies the following general assumption: there exists an $\mathscr{L}\in \mathcal{K}_{\infty}$ such that $\Vert h(x)-h(x')\Vert \leq \mathscr{L}(\Vert x-x'\Vert)$ for all $x,x' \in X$. 
\begin{remark}\label{Remark 1}
	Note that our assumption on $h$ with $\mathscr{L}\in \mathcal{K}_{\infty}$ is more general than the standard Lipschitz condition in which $\mathscr{L}$ is a linear function (i.e., $\mathscr{L}(\alpha) = L\alpha$ for some nonnegative $L$). Moreover, this assumption on $h$ is not restrictive provided that $h$ is continuous and one works on a compact subset of $X$. More precisely, all uniformly continuous functions automatically satisfy this assumption \cite{2003math}.
\end{remark}

Given the dt-SS in \eqref{eq:dt-SS}, we are interested in \emph{Markov policies} to control the system defined as follows.
\begin{definition}
	A Markov policy for the dt-SS $\Sigma$ in \eqref{eq:dt-SS} is a sequence
	$\rho = (\rho_0,\rho_1,\rho_2,\ldots)$ of universally measurable stochastic kernels $\rho_n$ \cite{BS96},
	each defined on $P = \{1,\dots, m \}$, given $X\times W$. The class of all such Markov policies is denoted by $\Pi_M$.
\end{definition} 

In this paper, we are interested in studying interconnected dt-SS without internal inputs that results from the interconnection of dt-SS having both internal inputs and switching signals. In this case, the interconnected dt-SS without internal inputs is indicated by the simplified tuple $(X,P, \mathcal{P},\varsigma, F, Y, h)$ with $f_p:X\times \mathcal{V}_\varsigma\rightarrow X$, $\forall p\in P$. 

\subsection{Global Markov Decision Processes}\label{subsec:MDP}
A dt-SS $\Sigma$ in \eqref{eq:dt-SS} can be \emph{equivalently} reformulated as an infinite Markov decision process (MDP)~\cite[Proposition 7.6, pp. 122]{kallenberg1997foundations}
\begin{equation}
\Sigma=(X,P,\mathcal{P},W,T_{\mathsf x},Y,h),	
\end{equation}
where $T_{\mathsf x}:\mathcal B(X)\times X\times P\times W\rightarrow[0,1]$,
is a conditional stochastic kernel that assigns to any $x \in X$, $ p\in P$, and $w\in W$ a probability measure $T_{\mathsf x}(\cdot | x,p, w)$
on the measurable space
$(X,\mathcal B(X))$
so that for any set $\mathcal{A} \in \mathcal B(X)$, 
$$\mathds{P} (x(k+1)\in \mathcal{A}| x(k),\bold{p}(k),w(k)) = \int_\mathcal{A} T_{\mathsf x} (d x'|x(k),\bold{p}(k),w(k)).$$

For given $\bold{p}(\cdot), w(\cdot),$  the stochastic kernel $T_{\mathsf x}$ captures the evolution of the state of $\Sigma$ and can be uniquely determined by the pair $(\varsigma,f)$ using \eqref{Eq_1a}.

In this paper, we consider $\Sigma_p, \forall p\in P$, as \emph{local} MDPs and introduce the notion of \emph{global} Markov decision processes as in the next definition. 
Note that this notion is adapted from the definition of labeled transition systems defined in~\cite{baier2008principles} and modified to capture the stochastic nature of the system. This notion provides an alternative description of switched systems enabling us to represent a switched system and its finite MDP in a common framework. 

\begin{definition}\label{def: gMDP}
	Given a dt-SS $\Sigma=(X,P,\mathcal{P},W,\varsigma, F, Y, h)$, we define the associated global MDP $\mathbb{G}(\Sigma) = (\mathbb{X},\mathbb{U},\mathbb{W},\varsigma,\mathbb{F},\mathbb{Y},\mathbb{H})$,
	where:
	\begin{itemize}
		\item $\mathbb{X} = X \times P \times \{0,\dots,k_d-1\}$ is the set of states. A state $(x,p,l) \in \mathbb{X}$ means that the current state of $\Sigma$ is $x$, the current value of the switching signal is $p$, and the time elapsed since the latest switching time instant capped by $k_d$ is $l$;
		\item $\mathbb{U} = P$ is the set of \emph{external} inputs;
		\item $\mathbb{W}=W$ is the set of \emph{internal} inputs;
		\item $\varsigma$ is a sequence of i.i.d. random variables;
		\item $\mathbb{F}:\mathbb{X}\times \mathbb{U}\times\mathbb{W}\times \mathcal{V}_{\varsigma} \rightarrow \mathbb{X}$\, is the one-step transition function given by $(x',p',l')=\mathbb{F}\,((x,p,l),\nu,w,\varsigma)$ if and only if $x'=f_p(x,w,\varsigma)$, $\nu=p$ and the following scenarios hold:
		\begin{itemize}
			\item $l<k_d-1, p'=p$, and $l' = l+1$: switching is not allowed because the time elapsed since the latest switch is strictly smaller than the dwell-time;
			\item $l=k_d-1, p'=p$, and $l'=k_d-1$: switching is allowed but no switch occurs;
			\item $l=k_d-1, p'\neq p$, and $l' = 0$: switching is allowed and a switch occurs;
		\end{itemize}	
		\item $\mathbb{Y} = Y$ is the output space;
		\item $\mathbb{H}:\mathbb{X} \rightarrow \mathbb{Y}$ is the output map defined as $\mathbb{H}\,(x,p,l) = h(x)$.
	\end{itemize}
	We associate respectively to $\mathbb{U}$ and $\mathbb{W}$ the sets $\mathcal U$ and $\mathcal W$ to be collections of sequences $\{\nu(k):\Omega\rightarrow \mathbb{U},\,\,k\in\N\}$ and $\{w(k):\Omega\rightarrow \mathbb{W},\,\,k\in\N\}$, in which $\nu(k)$ and $w(k)$ are independent of $\varsigma(t)$ for any $k,t\in\mathbb N$ and $t\ge k$. We also denote the initial conditions of $p$ and $l$ by $p_0$ and $l_0=0$.
	
	\begin{remark}
		Note that in the global MDP $\mathbb{G}(\Sigma)$ in Definition~\ref{def: gMDP}, we added two additional variables $p$ and $l$ to the state tuple of the system $\Sigma$, in which $l$ is a counter that depending on its value allows or prevents the system from switching, and $p$ acts as a memory to record the input.
	\end{remark}
	
	\begin{remark}
		Note that we employ the term ``internal'' for inputs and outputs of subsystems that are affecting each other in the interconnection topology: an internal output of a subsystem affects an internal input of another subsystem. We utilize the term ``external'' for inputs and outputs that are not employed for the sake of constructing the interconnection. Properties of the interconnected system are specified over external outputs. The main goal is to synthesize external inputs (switching signals) to satisfy desired properties over external outputs. 
	\end{remark}
\end{definition}

\begin{proposition}\label{Proposition}
	Global MDP $\mathbb{G}(\Sigma)$ in Definition~\ref{def: gMDP} is itself an MDP and the output trajectory of $\Sigma$ defined in~\eqref{Eq_1a} can be uniquely mapped to an output trajectory of $\mathbb{G}(\Sigma)$ and vice versa.
\end{proposition}

The proof of Proposition~\ref{Proposition} is provided in the Appendix.

\subsection{Finite Markov Decision Processes}

In this subsection, we approximate a dt-SS $\Sigma$ with a \emph{finite} $\widehat\Sigma$ using Algorithm~\ref{algo:MC_app}. 
To construct such a finite approximation, we assume the state and input sets of the \mbox{dt-SS} $\Sigma$ are restricted to compact subsets over which we are interested to perform synthesis. The rest of the state sets can be considered as single absorbing states in both $\Sigma$ and $\widehat\Sigma$. In order to make the notation easier, we assume this procedure is already applied to the system and eliminate the absorbing states from the presentation.

Algorithm~\ref{algo:MC_app} first constructs a finite partition from the state set $X$ and internal input set $W$.
Then representative points $\bar x_i\in \mathsf X_i$, and $\bar w_i\in \mathsf W_i$ are selected as abstract states and internal inputs.
Transition probabilities in the finite MDP $\widehat\Sigma$ are also computed according to \eqref{eq:trans_prob}. The output map $\hat h$ is the same as $h$ with its domain restricted to finite state set $\hat X$ (cf. Step \ref{step:output_map}) and the output set $\hat Y$ is the image of $\hat X$ under $h$ (cf. Step \ref{step:output_space}).

\begin{algorithm}[h]
	\caption{Abstraction of dt-SS $\Sigma$ by a finite MDP $\widehat\Sigma$}
	\label{algo:MC_app}
	\begin{center}
		\begin{algorithmic}[1]
			\REQUIRE 
			Input dt-SS $\Sigma=(X,P, \mathcal{P},W,T_{\mathsf x},Y,h)$
			\STATE
			Select finite partitions of sets $X,W$ as $X = \cup_{i=1}^{n_x} \mathsf X_i$, $W = \cup_{i=1}^{n_w} \mathsf W_i$				
			\STATE
			For each $\mathsf X_i$, and $\mathsf W_i$, select single representative points $\bar x_i \in \mathsf X_i$, $\bar w_i \in \mathsf W_i$
			\STATE
			Define 
			$\hat X := \{\bar x_i, i=1,...,n_x\}$ as the finite state set of MDP~$\widehat\Sigma$ with internal input set $\hat W := \{\bar w_i, i=1,...,n_w\}$
			\STATE
			\label{step:refined}
			Define the map $\Xi:X\rightarrow 2^X$ that assigns to any $x\in X$, the corresponding partition set it belongs to, i.e.,
			$\Xi(x) = \mathsf X_i$ if $x\in \mathsf X_i$ for some $i=1,2,\ldots,n_x$
			\STATE
			Compute the discrete transition probability matrix $\hat T_{\mathsf x}$ for $\widehat\Sigma$ as:
			\begin{equation}
			\label{eq:trans_prob}
			\hat T_{\mathsf x} (x'|x,p,w) 
			= T_{\mathsf x} (\Xi(x')|x,p,w),
			\end{equation}
			for all $x,x'\in \hat X, p\in P, w\in\hat W$
			\STATE
			\label{step:output_space}
			Define the output space $\hat Y := h(\hat X)$
			\STATE
			\label{step:output_map}
			Define the output map $\hat h := h|_{\hat X}$
			\ENSURE
			Finite MDP
			\begin{equation}
			\label{Finite MDPs}
			\widehat\Sigma = (\hat X,P, \mathcal{P},\hat W, \hat T_{\mathsf x}, \hat Y, \hat h)
			\end{equation}
		\end{algorithmic}
	\end{center}
\end{algorithm}

\begin{remark}\label{Def154}
	Given a dt-SS $\Sigma=(X,P, \mathcal{P}, W,\varsigma,F,Y,h)$ with $F = \{f_1,\dots, f_m \}$,
	the finite MDP $\widehat\Sigma$ constructed in Algorithm~\ref{algo:MC_app} can be represented as 
	\begin{equation}
	\label{eq:abs_tuple}
	\widehat\Sigma =(\hat X,P, \mathcal{P},\hat W, \varsigma,\hat F, \hat Y, \hat h),
	\end{equation}
	with $\hat F = \{\hat f_1,\dots, \hat f_m \}$, where $\hat f_p:\hat X\times\hat W\times \mathcal{V}_\varsigma\rightarrow\hat X, \forall p\in P,$ is defined as
	\begin{equation}\label{Abstraction Map}
	\hat f_p(\hat{x},\hat{w},\varsigma) = \Pi_{x}(f_p(\hat{x},\hat{w},\varsigma)),	
	\end{equation}
	and $\Pi_x:X\rightarrow \hat X$ is the map that assigns to any $x\in X$, the representative point $\bar x\in\hat X$ of the corresponding partition set containing $x$.
	The initial state of $\widehat\Sigma$ is also selected according to $\hat x_0 := \Pi_x(x_0)$ with $x_0$ being the initial state of $\Sigma$. 
\end{remark}

Dynamical representation provided by Remark~\ref{Def154} uses the map $\Pi_x:X\rightarrow \hat X$ that
satisfies the inequality
\begin{equation}
\label{eq:Pi_delta}
\Vert \Pi_x(x)-x\Vert \leq \bar\delta,\quad \forall x\in X,
\end{equation}
where $\bar \delta:=\sup\{\|x-x'\|,\,\, x,x'\in \mathsf X_i,\,i=1,2,\ldots,n_x\}$ is the state discretization parameter. Now we have all the ingredients to formally define the finite abstraction of global MDPs as in the following definition.

\begin{definition}\label{def: abstract gMDP}
	Given a global MDP $\mathbb{G}(\Sigma) = (\mathbb{X},\mathbb{U},\mathbb{W},\varsigma,\mathbb{F},\mathbb{Y},\mathbb{H})$ associated with $\Sigma$ as in the Definition \ref{def: gMDP}, one can construct its finite abstraction as a finite global MDP  $\mathbb{\widehat G}(\widehat \Sigma) = (\mathbb{\hat X},\mathbb{\hat U},\mathbb{\hat W},\varsigma,\mathbb{\hat F},\mathbb{\hat Y},\mathbb{\hat H})$, where:
	\begin{itemize}
		\item $\mathbb{\hat X} = \hat X \times P \times \{0,\dots,k_d-1\}$ is the set of states;
		\item $\mathbb{\hat U} = \mathbb{U} = P$ is the set of \emph{external} inputs that remains the same as in the global MDP;	
		\item $\mathbb{\hat W}= \hat W$ is the set of \emph{internal} inputs;	
		\item $\varsigma$ is a sequence of i.i.d. random variables;
		\item $\mathbb{\hat F}:\mathbb{\hat X}\times \mathbb{\hat U}\times\mathbb{\hat W}\times \mathcal{V}_{\varsigma} \rightarrow \mathbb{\hat X}$\, is the one-step transition function given by $(\hat x',p',l')=\mathbb{\hat F}\,((\hat x,p,l),\hat \nu,\hat w,\varsigma)$ if and only if $ \hat x' = \hat f_p(\hat{x},\hat{w},\varsigma)$ as defined in~\eqref{Abstraction Map}, $\hat \nu=p$ and the following scenarios hold: 
		\begin{itemize}
			\item $l<k_d-1$, $p' = p$, and $l' = l+1$;
			\item $l=k_d-1$, $p' = p$, and $l'=k_d-1$;
			\item $l=k_d-1, p'\neq p$, and $l' = 0$;
		\end{itemize}
		\item $\mathbb{\hat Y} = \{\mathbb{H}(\hat x,p,l)\,|\,(\hat x,p,l)\in\mathbb{\hat X}\}$ is the output set;
		\item $\mathbb{\hat H}:\mathbb{\hat X} \rightarrow \mathbb{\hat Y}$ is the output map defined as $\mathbb{\hat H}\,(\hat x,p,l) = \mathbb{H}\,(\hat x,p,l) = h(\hat x)$.
	\end{itemize}
\end{definition}	

In the next section, in order to provide an approach for \emph{compositional synthesis} of interconnected dt-SS, we define the notions of stochastic pseudo-simulation and simulation functions. These two notions are employed to quantify the probabilistic error between a global MDP and its finite abstraction and also their interconnection without internal inputs, respectively.

\section{Stochastic Pseudo-Simulation and Simulation Functions}
\label{sec:SPSF}
In this section, we first introduce a notion of stochastic pseudo-simulation functions for dt-SS with internal inputs. We then define a notion of stochastic simulation functions for switched systems without internal inputs. We employ these definitions mainly to quantify closeness of a global MDP and its finite abstraction.

\begin{definition}\label{Def_1a} 
	Consider two global MDPs $\mathbb{G}(\Sigma) = (\mathbb{X},\mathbb{U},\mathbb{W},\varsigma,\mathbb{F},\mathbb{Y},\mathbb{H})$ and
	$\mathbb{\widehat G}(\widehat \Sigma) = (\mathbb{\hat X},\mathbb{\hat U},\mathbb{\hat W},\varsigma,\mathbb{\hat F},\mathbb{\hat Y},\mathbb{\hat H})$. A function $V:\mathbb{X}\times\mathbb{\hat X}\to\R_{\ge0}$ is called a stochastic pseudo-simulation function (SPSF) from  $\mathbb{\widehat G}(\widehat \Sigma)$ to $\mathbb{G}(\Sigma)$ if there exist $\alpha\in\mathcal{K}_\infty$,~$0<\kappa<1$, $ \rho_{\mathrm{int}}\in\mathcal{K}_\infty\cup\{0\}$, and a constant $\psi \in\R_{\ge 0}$ such that
	\begin{itemize}
		\item $\forall(x,p,l)\in \mathbb{X}, \forall(\hat x,p,l)\in \mathbb{\hat X}$,
		\begin{align}\label{Eq_2a}
		\alpha(\Vert \mathbb{H}(x,p,l)-\mathbb{\hat H}(\hat x,p,l)\Vert)\le V((x,p,l),(\hat x,p,l)),
		\end{align}
		\item  $\forall(x,p,l)\in \mathbb{X}, \forall(\hat x,p,l)\in \mathbb{\hat X}$, $\forall \hat \nu\in \mathbb{\hat U}, \forall w\in \mathbb{W},\forall \hat w\in \mathbb{\hat W}$,
		\begin{align}\notag
		\mathbb{E} \Big[V&((x',p',l'),(\hat x',p',l'))\,\big|\,x,\hat{x},p,l,w,\hat w\Big]\\\label{Eq_3a}
		\leq& \max\Big\{\kappa V((x,p,l),(\hat x,p,l)), \rho_{\mathrm{int}}(\Vert w-\hat w\Vert),\psi\Big\},
		\end{align}
		where the expectation operator $\mathbb E$ is with respect to $\varsigma$ under the one-step transition of both global MDPs with $\nu = \hat\nu$, i.e., $(x',p',l')=\mathbb{F}\,((x,p,l),\hat \nu,w,\varsigma)$ and $(\hat x',p',l')=\mathbb{\hat F}\,((\hat x,p,l),\hat \nu,\hat w,\varsigma)$.
	\end{itemize}
\end{definition}

If there exists an SPSF $V$ from $\mathbb{\widehat G}(\widehat \Sigma)$ to $\mathbb{G}(\Sigma)$, this is denoted by $\mathbb{\widehat G}(\widehat \Sigma)\preceq_{\mathcal{PS}}\mathbb{G}(\Sigma)$, and the system $\mathbb{\widehat G}(\widehat \Sigma)$ is called an abstraction of concrete (original) global MDP $\mathbb{G}(\Sigma)$.

Now, we modify the above notion for global MDPs without internal inputs by eliminating all the terms related to $w,\hat w$ which will be employed later for relating interconnected systems.

\begin{definition}\label{Def_2a} 
	Consider two global MDPs $\mathbb{G}(\Sigma) = (\mathbb{X},\mathbb{U},\varsigma,\mathbb{F},\mathbb{Y},\mathbb{H})$ and
	$\mathbb{\widehat G}(\widehat \Sigma) = (\mathbb{\hat X},\mathbb{\hat U},\varsigma,\mathbb{\hat F},\mathbb{\hat Y},\mathbb{\hat H})$ without internal inputs.
	A function $V:\mathbb{X}\times\mathbb{\hat X}\to\R_{\ge0}$ is called a \emph{stochastic simulation function} (SSF) from $\mathbb{\widehat G}(\widehat \Sigma)$ to $\mathbb{G}(\Sigma)$ if
	\begin{itemize}
		\item there exists $\alpha\in \mathcal{K}_{\infty}$ such that $\forall(x,p,l)\in \mathbb{X}, \forall(\hat x,p,l)\in \mathbb{\hat X}$,
		\begin{align}\label{eq:lowerbound2}
		\alpha(\Vert \mathbb{H}(x,p,l)-\mathbb{\hat H}(\hat x,p,l)\Vert)\le V((x,p,l),(\hat x,p,l)),
		\end{align}
		\item  $\forall(x,p,l)\in \mathbb{X}, \forall(\hat x,p,l)\in \mathbb{\hat X}$, $\forall \hat \nu\in \mathbb{\hat U}$,
		\begin{align}\label{eq6666}
		\mathbb{E} \Big[&V((x',p',l'),(\hat x',p',l'))\,\big|\,x,\hat{x},p,l\Big]\leq \max\Big\{\kappa V((x,p,l),(\hat x,p,l)),\psi\Big\},\
		\end{align}
		for some $0<\kappa<1$, and $\psi \in\R_{\ge 0}$, where the expectation operator $\mathbb E$ is with respect to $\varsigma$ under the one-step transition of both global MDPs with $\nu = \hat\nu$, i.e., $(x',p',l')=\mathbb{F}\,((x,p,l),\hat \nu,\varsigma)$ and $(\hat x',p',l')=\mathbb{\hat F}\,((\hat x,p,l),\hat \nu,\varsigma)$.
	\end{itemize}
\end{definition}
If there exists an SSF $V$ from $\mathbb{\widehat G}(\widehat \Sigma)$ to $\mathbb{G}(\Sigma)$, this is denoted by $\mathbb{\widehat G}(\widehat \Sigma)\preceq \mathbb{G}(\Sigma)$, and $\mathbb{\widehat G}(\widehat \Sigma)$ is called an abstraction of $\mathbb{G}(\Sigma)$.

\begin{remark}
	Note that conditions~\eqref{Eq_2a}, \eqref{Eq_3a}, \eqref{eq:lowerbound2}, and \eqref{eq6666} roughly speaking guarantee that if the concrete system and its abstraction start from two close initial conditions, then their outputs remain close (in terms of expectation) after one step. This type of conditions is closely related to the ones in the notions of (bi)simulation relations~\cite{tabuada2009verification}.
\end{remark}

In order to show the usefulness of SSF in comparing output trajectories of two global MDPs (without internal inputs) in a probabilistic setting, we need the following technical lemma borrowed from~\cite[Theorem 3, pp. 86]{1967stochastic} with some slight modifications adapted to stochastic switched systems.

\begin{lemma}\label{Lemma: Kushner}
	Let $\mathbb{G}(\Sigma) = (\mathbb{X},\mathbb{U},\varsigma,\mathbb{F},\mathbb{Y},\mathbb{H})$ be a global MDP with the transition map $\mathbb{F}:\mathbb{X}\times \mathbb{U}\times \mathcal{V}_{\varsigma} \rightarrow \mathbb{X}$. Assume there exist $V:\mathbb{X}\to\R_{\ge0}$ and constants $0<\kappa<1$, and $\psi \in\R_{\ge 0}$ such that		
	\begin{align}\notag
	\mathbb{E} \Big[V(x',p',l')\,\big|\,x,p,l\Big]\leq \kappa V(x,p,l)+\psi,
	\end{align}	
	where $(x',p',l')=\mathbb{F}\,((x,p,l),p,\varsigma)$.
	Then for any random variable $a$ as the initial state of the underlying dt-SS, any initial mode $p_0$, and $l_0=0$ as the initial counter, the following inequity holds:
	\begin{equation}\notag
	\PP\left\{\sup_{0\leq k\leq T_d} V(x(k),p(k),l(k))\geq\varepsilon\,|\,a,p_0\right\}\le\delta,
	\end{equation}
	\begin{equation*}
	\delta := 
	\begin{cases}
	1-(1-\frac{V(a,p_0,l_0)}{\varepsilon})(1-\frac{\psi}{\varepsilon})^{T_d}, & \quad\quad\text{if}~\varepsilon\geq\frac{\psi}{\kappa},\\
	(\frac{V(a,p_0,l_0)}{\varepsilon})(1-\kappa)^{T_d}+(\frac{\psi}{\kappa\varepsilon})(1-(1-\kappa)^{T_d}), & \quad\quad\text{if}~\varepsilon<\frac{\psi}{\kappa}.
	\end{cases}
	\end{equation*}
\end{lemma}

Now by employing Lemma~\ref{Lemma: Kushner}, we provide one of the results of the paper.

\begin{theorem}\label{Thm_1a}
	Let $\mathbb{G}(\Sigma) = (\mathbb{X},\mathbb{U},\varsigma,\mathbb{F},\mathbb{Y},\mathbb{H})$ and
	$\mathbb{\widehat G}(\widehat \Sigma) = (\mathbb{\hat X},\mathbb{\hat U},\varsigma,\mathbb{\hat F},\mathbb{\hat Y},\mathbb{\hat H})$ be two global MDPs without internal inputs. Suppose $V$ is an SSF from $\mathbb{\widehat G}(\widehat \Sigma)$ to $\mathbb{G}(\Sigma)$. For any random variables $a$ and $\hat a$ as the initial states of the two dt-SS, any initial mode $p_0$, and for any external input trajectory $\hat\nu(\cdot)\in\mathcal{\hat U}$ that preserves the Markov property for the closed-loop $\mathbb{\widehat G}(\widehat \Sigma)$, the following inequality holds:
	\begin{align}\label{Eq_25}
   \mathds{P}&\left\{\sup_{0\leq k\leq T_d}\Vert y_{a\hat\nu}(k)-\hat y_{\hat a\hat\nu }(k)\Vert\geq\varepsilon\,|\,a,\hat a,p_0\right\}\\\notag
	&\leq
	\begin{cases}
	1-(1-\frac{V((a,p_0,l_0),(\hat a, p_0,l_0))}{\alpha\left(\varepsilon\right)})(1-\frac{\psi}{\alpha\left(\varepsilon\right)})^{T_d}, & \quad\quad\text{if}~\alpha\left(\varepsilon\right)\geq\frac{\psi}{\kappa},\\
	\frac{V((a,p_0,l_0),(\hat a, p_0,l_0))}{\alpha\left(\varepsilon\right)}(1-\kappa)^{T_d}+\frac{\psi}{\kappa\alpha\left(\varepsilon\right)}(1-(1-\kappa)^{T_d}), & \quad\quad\text{if}~\alpha\left(\varepsilon\right)<\frac{\psi}{\kappa}.
	\end{cases}
	\end{align}
\end{theorem}
The proof of Theorem~\ref{Thm_1a} is provided in the Appendix.

\section{Compositional Abstractions for Interconnected Switched Systems}
\label{sec:compositionality}
In this section, we analyze networks of stochastic switched subsystems by driving a small-gain type
condition and discuss how to construct their \emph{finite} global MDP together with a simulation function based on the corresponding SPSF of their subsystems. 

\subsection{Concrete Interconnected Stochastic Switched Systems}
Suppose we are given $N$ \emph{concrete} stochastic switched subsystems
\begin{equation}
\label{eq:network}
\Sigma_i=(X_i,P_i, \mathcal{P}_i,W_i,\varsigma_i, F_i, Y_i, h_i),~i\in \{1,\dots,N\},
\end{equation}
with its \emph{equivalent} global MDP $\mathbb{G}(\Sigma_i) = (\mathbb{X}_i,\mathbb{U}_i,\mathbb{W}_i,\varsigma_i,\mathbb{F}_i,\mathbb{Y}_i,\mathbb{H}_i)$, in which their internal inputs and outputs are partitioned as
\begin{align}\label{config1}
w_i=[{w_{i1};\ldots;w_{i(i-1)};w_{i(i+1)};\ldots;w_{iN}}],\quad\quad y_i=[{y_{i1};\ldots;y_{iN}}],
\end{align}
and their output spaces and functions are of the form
\begin{align}\label{config2}
Y_i=\prod_{j=1}^{N}Y_{ij},\quad h_{i}(x_i)=[h_{i1}(x_i);\ldots;h_{iN}(x_i)].
\end{align}
We interpret the outputs $y_{ii}$ as \emph{external} ones, whereas the outputs $y_{ij}$ with $i\neq j$ are \emph{internal} ones which are utilized to interconnect stochastic switched subsystems. For the interconnection, we assume that $w_{ij}$ is equal to $y_{ji}$ if there is a connection from $\Sigma_{j}$ to $\Sigma_i$, otherwise we put the connecting output function identically zero, i.e. $h_{ji}\equiv 0$. 
Now, we are ready to define the \emph{interconnection} of concrete dt-SS $\Sigma_i=(X_i,P_i, \mathcal{P}_i,W_i,\varsigma_i, F_i, Y_i, h_i)$.
\begin{definition}
	Consider $N\in\mathbb N_{\geq1}$ dt-SS $\Sigma_i=(X_i,P_i, \mathcal{P}_i,W_i,\varsigma_i, F_i, Y_i, h_i)$, with the input-output configuration as in \eqref{config1} and \eqref{config2}. The interconnection of  $\Sigma_i$, $\forall i\in \{1,\ldots,N\}$, is the \emph{concrete} interconnected dt-SS $\Sigma = (X,P, \mathcal{P},\varsigma, F, Y, h)$, denoted by
	$\mathcal{I}(\Sigma_1,\ldots,\Sigma_N)$, such that $X:=\prod_{i=1}^{N}X_i$,  $P:=\prod_{i=1}^{N}P_i$, $\mathcal{P}:=\prod_{i=1}^{N}\mathcal{P}_i$, $F:=\prod_{i=1}^{N}F_{i}$, $Y:=\prod_{i=1}^{N}Y_{ii}$, and $h=\prod_{i=1}^{N}h_{ii}$, subjected to the following constraint:
	\begin{equation*}
	\forall i,j\in \{1,\dots,N\},i\neq j\!: ~~~ w_{ji} = y_{ij}, \quad\quad Y_{ij}\subseteq W_{ji}.
	\end{equation*}
	
\end{definition}
Similarly, given global MDPs $\mathbb{G}(\Sigma_i) = (\mathbb{X}_i,\mathbb{U}_i,\mathbb{W}_i,\varsigma_i,\mathbb{F}_i,\mathbb{Y}_i,\mathbb{H}_i),i\in \{1,\dots,N\}$, one can also define the \emph{interconnection} of concrete global MDPs $\mathbb{G}(\Sigma_i)$ as 	$\mathcal{I}(\mathbb{G}(\Sigma_1),\ldots,\mathbb{G}(\Sigma_N))$.

Now assume that any \emph{concrete} global MDP $\mathbb{G}(\Sigma_i) = (\mathbb{X}_i,\mathbb{U}_i,\mathbb{W}_i,\varsigma_i,\mathbb{F}_i,\mathbb{Y}_i,\mathbb{H}_i),i\in \{1,\dots,N\}$, admits an \emph{abstract} global MDP $\mathbb{\widehat G}(\widehat \Sigma_i) = (\mathbb{\hat X}_i,\mathbb{\hat U}_i,\mathbb{\hat W}_i,\varsigma_i,\mathbb{\hat F}_i,\mathbb{\hat Y}_i,\mathbb{\hat H}_i)$ together with an SPSF $V_i$ from $\mathbb{\widehat G}(\widehat \Sigma_i)$ to $\mathbb{G}(\Sigma_i)$ with the corresponding functions and constants denoted by $\alpha_i, \rho_{\mathrm{int}i},\kappa_i$ and $\psi_i$ as in Definition~\ref{Def_1a}.

\subsection{Compositional Abstractions of Interconnected Switched Systems}
In order to provide compositionality results of the paper, we first define the abstraction map $\Pi_{w_{ji}}$ on $W_{ji}$ that assigns to any $w_{ji}\in W_{ji}$, a representative point $\bar w_{ji}\in \hat W_{ji}$ of the corresponding partition set containing $w_{ji}$. The mentioned map satisfies 
\begin{equation}
\label{eq:Pi_mu}
\Vert \Pi_{w_{ji}}(w_{ji})-w_{ji} \Vert \leq \bar \mu_{ji},\,\quad \forall w_{ji}\in W_{ji},
\end{equation}	
where $\bar \mu_{ji}$ is an \emph{internal input} discretization parameter defined similar to $\bar \delta$ in~\eqref{eq:Pi_delta}. 

\begin{remark}
	Note that condition~\eqref{eq:Pi_mu} helps us to choose quantization parameters of internal input sets freely at the cost of incurring an additional error term formulated in $\psi$ in~\eqref{overall-error}.
\end{remark}

Now, we define a notion of the \emph{interconnection} of \emph{abstract} global MDPs $\mathbb{\widehat G}(\widehat \Sigma_i) = (\mathbb{\hat X}_i,\mathbb{\hat U}_i,\mathbb{\hat W}_i,\varsigma_i,\mathbb{\hat F}_i,\mathbb{\hat Y}_i,\mathbb{\hat H}_i)$.

\begin{definition}
	Consider $N\in\mathbb N_{\geq1}$ \emph{abstract} global MDPs $\mathbb{\widehat G}(\widehat \Sigma_i) = (\mathbb{\hat X}_i,\mathbb{\hat U}_i,\mathbb{\hat W}_i,\varsigma_i,\mathbb{\hat F}_i,\mathbb{\hat Y}_i,\mathbb{\hat H}_i)$, with the input-output configuration similar to~\eqref{config1} and \eqref{config2}. The interconnection of  $\mathbb{\widehat G}(\widehat\Sigma_i)$, $\forall i\in \{1,\ldots,N\}$, is the interconnected \emph{abstract}  global MDP $\mathbb{\widehat G}(\widehat \Sigma) = (\mathbb{\hat X},\mathbb{\hat U},\varsigma,\mathbb{\hat F},\mathbb{\hat Y},\mathbb{\hat H})$, denoted by
	$\mathcal{\widehat I}(\mathbb{\widehat G}(\widehat \Sigma_1),\ldots,\mathbb{\widehat G}(\widehat \Sigma_N))$, such that $\mathbb{\hat X}:=\prod_{i=1}^{N}\mathbb{\hat X}_i$,  $\mathbb{\hat U}:=\prod_{i=1}^{N}\mathbb{\hat U}_i$, $\mathbb{\hat Y}:=\prod_{i=1}^{N} \mathbb{\hat Y}_{ii}$, $\mathbb{\hat H}=\prod_{i=1}^{N}\mathbb{\hat H}_{ii}$, and the map $\mathbb{\hat F}=\prod_{i=1}^{N}\mathbb{\hat F}_{i}$  is the transition function given by $(\hat x',p',l')=\mathbb{\hat F}\,((\hat x,p,l),\hat \nu,\hat w,\varsigma)$ if and only if $ \hat x' = \hat f_p(\hat{x},\hat{w},\varsigma)$ as defined in~\eqref{Abstraction Map}, $\hat \nu=p$ and the following scenarios hold for any $i\in\{1,\dots,N\}$: 
	\begin{itemize}
		\item $l_i<k_{d_i}-1$, $p'_i = p_i$, and $l'_i = l_i+1$;
		\item $l_i=k_{d_i}-1$, $p'_i = p_i$, and $l'_i=k_{d_i}-1$;
		\item $l_i=k_{d_i}-1$, $p'_i\neq p_i$, and $l'_i = 0$;
	\end{itemize}where $\hat x= [\hat x_1;\dots;\hat x_N], \hat \nu= [\hat \nu_1;\dots;\hat \nu_N], p= [p_1;\dots;p_N], l= [l_1;\dots;l_N]$, and subjected to the following constraint:
	\begin{align}\notag
	\forall i,j\in \{1,\dots,N\},i\neq j\!: \hat w_{ji} = \Pi_{w_{ji}}(\hat y_{ij}), \quad\quad\Pi_{w_{ji}}(\mathbb{\hat Y}_{ij})\subseteq\mathbb{\hat W}_{ji}.
	\end{align}
\end{definition}
Now we raise the following small-gain assumption inspired by the corresponding one in~\cite{dashkovskiy2007iss,dashkovskiy2010small} to establish the main compositionality results of the paper.

\begin{assumption}~\label{Assump: Gamma1}
	Assume that there exist $\mathcal{K}_\infty$ functions $\tilde \delta_f, \bar \lambda$ such that $(\bar \lambda - \mathcal{I}_d)\in\mathcal{K}_\infty$ and $\mathcal{K}_\infty$ functions $\kappa_{ij}$ defined as
	\begin{equation*}
	\kappa_{ij}(s) := 
	\begin{cases}
	\kappa_is ~~~~& \text{if }i = j,\\
	(\mathcal{I}_d + \tilde \delta_f)\circ\rho_{\mathrm{int}i}\circ \bar \lambda \circ\alpha_j^{-1}(s) ~~~~& \text{if }i \neq j,
	\end{cases}
	\end{equation*}
	satisfying
	\begin{equation}
	\label{Assump: Kappa1}
	\kappa_{i_1i_2}\circ\kappa_{i_2i_3}\circ\dots \circ \kappa_{i_{r-1}i_{r}}\circ\kappa_{i_{r}i_1} < \mathcal{I}_d
	\end{equation}	
	for all sequences $(i_1,\dots,i_{r}) \in \{1,\dots,N\}^{r}$ and ${r}\in \{1,\dots,N\}$.
	
	The small-gain condition~\eqref{Assump: Kappa1} implies the existence of $\mathcal{K}_\infty$ functions $\sigma_i>0$ \cite[Theorem 5.5]{ruffer2010monotone}, satisfying
	\begin{align}\label{compositionality1}
	\max_{i,j}\Big\{\sigma_i^{-1}\circ\kappa_{ij}\circ\sigma_j\Big\} < \mathcal{I}_d, ~~~~i,j = \{1,\dots,N\}.
	\end{align}
\end{assumption}	
In the next theorem, we leverage small-gain Assumption \ref{Assump: Gamma1} to quantify the error between the interconnection of concrete global MDPs and that of their finite abstractions in a compositional manner.

\begin{theorem}\label{Thm: Comp1}
	Consider the interconnected global MDP $\mathbb{G}(\Sigma) = (\mathbb{X},\mathbb{U},\varsigma,\mathbb{F},\mathbb{Y},\mathbb{H})$ induced by $N\in\mathbb N_{\geq1}$ global MDPs~$\mathbb{G}(\Sigma_i)$. Suppose that each $\mathbb{G}(\Sigma_i)$ admits a finite abstraction $\mathbb{\widehat  G}(\widehat \Sigma_i)$ together with an SPSF $V_i$. If Assumption~\ref{Assump: Gamma1} holds,
	then function $V((x,p,l),(\hat x,p,l))$ defined as
	\begin{equation}
	\label{Comp: Simulation Function1}
	V((x,p,l),(\hat x,p,l)) := \max_{i} \{\sigma_i^{-1}(V_i((x_i,p_i,l_i),(\hat x_i,p_i,l_i)))\},
	\end{equation}
	for $\sigma_i$ as in \eqref{compositionality1}, is an SSF function from $\mathcal{\widehat I}(\mathbb{\widehat G}(\widehat \Sigma_1),\ldots,\mathbb{\widehat G}(\widehat \Sigma_N))$ to $\mathcal{I}(\mathbb{G}(\Sigma_1),\ldots,\mathbb{G}(\Sigma_N))$ provided that $\max_{i}\sigma_i^{-1}$ is concave. 	
\end{theorem}

The proof of Theorem~\ref{Thm: Comp1} is provided in the Appendix.

\section{Construction of Stochastic Pseudo-Simulation Functions}
\label{sec:constrcution_finite}

In this section, we impose conditions on the \emph{concrete} dt-SS $\Sigma$ enabling us to find an SPSF from finite abstraction $\mathbb{\widehat G}(\widehat \Sigma)$ to $\mathbb{G}(\Sigma)$. The required conditions are first presented in a \emph{general setting} for nonlinear stochastic switched systems in Subsection~\ref{subsec:nonlinear} and then represented via some matrix inequalities for a class of \emph{nonlinear} stochastic switched systems in Subsection~\ref{subsec:linear}.

\subsection{General Setting of Nonlinear Stochastic Switched Systems}
\label{subsec:nonlinear}

The stochastic pseudo-simulation function from finite global MDP $\mathbb{\widehat G}(\widehat \Sigma)$ to $\mathbb{G}(\Sigma)$ is established under the assumption that the original discrete-time stochastic switched subsystems $\Sigma_p, \forall p\in P,$ are \emph{incrementally input-to-state stable} ($\delta$-ISS) as in the following definition.

\begin{definition}\label{Def11}
	A dt-SS  $\Sigma_p$ is called \emph{incrementally input-to-state stable} ($\delta$-ISS) if there exists function $V_p: X\times X\to \mathbb{R}_{\geq0} $  such that $\forall x, x'\in X$, $\forall w, w' \in W$, the following two inequalities hold:
	
	\begin{align}\label{Con55}
	\underline{\alpha}_p(\Vert x-x'\Vert ) \leq V_p(x,x')\leq \overline{\alpha}_p (\Vert x-x'\Vert ),
	\end{align}	
	and	
	\begin{align}\label{Con85}
	\mathbb{E} \Big[V_p(f_p(x,w,\varsigma),f_p(x',w',\varsigma))\,\big|\,x,x',w, w'\Big]\leq\bar{\kappa}_pV_p(x,x')+\bar \rho_{\mathrm{int}p}(\Vert w-w'\Vert),
	\end{align}
	for some $\underline{\alpha}_p, \overline{\alpha}_p \in \mathcal{K}_{\infty}$, $0<\bar{\kappa}_p<1$, and $\bar \rho_{\mathrm{int}p}\in\mathcal{K}_\infty\cup\{0\}$.
\end{definition}
The above definition is a stochastic counterpart of the $\delta$-ISS Lyapunov functions defined for discrete-time deterministic systems in~\cite{tran2016convergence}. In order to construct a stochastic pseudo-simulation function from finite global MDP $\mathbb{\widehat G}(\widehat \Sigma)$ to $\mathbb{G}(\Sigma)$, we need to raise the following assumptions. These assumptions are essential to show the main result of this section in Theorem~\ref{Thm_5a}.

\begin{assumption}\label{Assume: Switching}
	There exists $\mu\ge 1$ such that
	\begin{align}
	\forall x,x' \in X,~ \forall p,p' \in P, \quad V_p(x,x')\leq \mu V_{p'}(x,x').
	\end{align}
	
	\begin{remark}\label{existance mu}
		Assumption~\ref{Assume: Switching} is a standard one in switched systems accepting \emph{multiple} Lyapunov functions with \emph{dwell-time} similar to the one appeared in~\cite[equation (3.6)]{liberzon2003switching}.  Note that if function $V_p$ is quadratic in the form of~\eqref{Eq_77a}, there always exists $\mu\ge 1$ satisfying Assumption~\ref{Assume: Switching} as $\mu = \max(\frac{\lambda_{\max}(M_p)}{\lambda_{\min}(M_{p'})},\frac{\lambda_{\max}(M_{p'})}{\lambda_{\min}(M_p)}), \forall p,p' \in P$
		(cf. the second case study). If there exists a common Lyapunov function between all modes, then $\mu = 1$ and $V((x,p,l),(\hat x,p,l)) = V(x,\hat x)$ (cf. the first case study).
	\end{remark}
\end{assumption}

\begin{assumption}\label{Assume: Switching1}
	Assume that $\forall p \in P$, there exists a function $\gamma_p\in\mathcal{K}_{\infty}$  such that
	\begin{equation}\label{Eq65}
	V_p(x,x')-V_p(x,x'')\leq \gamma_p(\Vert x'-x''\Vert),\quad \forall x,x',x'' \in X.
	\end{equation}
\end{assumption}

\begin{remark}
	As shown in \cite{zamani2014symbolic} and by employing the mean value theorem, inequality \eqref{Eq65} is always satisfied for any differentiable function $V_p$ restricted to a compact subset of $X \times X$. Note that if one chooses $V_p=((x-x')^TM_p(x-x'))^{\tfrac{1}{2}}, \forall x,x' \in X$, then $\gamma_p(s)=\sqrt{\lambda_{\max}(M_p)}s, \forall s\in\mathbb R_{\ge0}$.
\end{remark}

Under Definition~\ref{Def11} and Assumptions~\ref{Assume: Switching} and~\ref{Assume: Switching1}, the next theorem shows a relation between $\mathbb{G}(\Sigma)$  and $\mathbb{\widehat G}(\widehat \Sigma)$ via establishing a stochastic pseudo-simulation function between them.
\begin{theorem}\label{Thm_5a}
	Let $\Sigma=(X,P, \mathcal{P},W,\varsigma, F, Y, h)$ be a switched system with its equivalent global MDP $\mathbb{G}(\Sigma) = (\mathbb{X},\mathbb{U},\mathbb{W},\varsigma,\mathbb{F},\mathbb{Y},\mathbb{H})$. Consider an abstract global MDP $\mathbb{\widehat G}(\widehat \Sigma) = (\mathbb{\hat X},\mathbb{\hat U},\mathbb{\hat W},\varsigma,\mathbb{\hat F},\mathbb{\hat Y},\mathbb{\hat H})$ constructed as in Definition \ref{def: abstract gMDP}. For any $p\in P$, let $\Sigma_p$ be an \emph{incrementally input-to-state stable} ($\delta$-ISS) dt-SS  via a
	function $V_p$ as in Definition~\ref{Def11}, and Assumptions~\ref{Assume: Switching} and~\ref{Assume: Switching1} hold. Let $\epsilon>1$. If \,$\forall p\in P$, $~k_d \geq \epsilon\frac{\ln(\mu)}{\ln(1/\bar\kappa_p)}+1$, then 
	\begin{align}\label{function V}
	V((x,p,l),(\hat x,p,l))=\frac{1}{{\bar\kappa_p}^{l/\epsilon}}V_p(x,\hat x),
	\end{align} 
	is an SPSF from $\mathbb{\widehat G}(\widehat \Sigma)$ to $\mathbb{G}(\Sigma)$.
\end{theorem}
The proof of Theorem~\ref{Thm_5a} is provided in the Appendix.

\begin{remark}
	Note that if there exists a common Lyapunov function $V: X\times X\to\R_{\ge0}$ between all switching modes $p\in P$ satisfying Definition~\ref{Def11} and Assumptions~\ref{Assume: Switching} (with $\mu = 1$) and~\ref{Assume: Switching1}, then $V((x,p,l),(\hat x,p,l)) = V(x,\hat x)$ and Definitions~\ref{Def_1a} and \ref{Def_2a} reduce to Definitions $3.1$ and $9.1$ in~\cite{lavaei2018ADHSJJ} (cf. the first case study).
\end{remark}

Now we provide similar results as in Subsection \ref{subsec:nonlinear} but tailored to a particular class of \emph{nonlinear} stochastic switched systems.

\subsection{Stochastic Switched Systems with Slope Restrictions on Nonlinearity}
\label{subsec:linear}
Here, we focus on a specific class of discrete-time \emph{nonlinear} stochastic switched systems $\Sigma$ together with \emph{quadratic} functions $V_p$ and provide an approach on the construction of their finite global MDPs. The class of nonlinear switched systems is given by
\begin{align}\label{Eq_58a}
\Sigma:\left\{\hspace{-1.5mm}\begin{array}{l}x(k+1)=A_{\bold{p}(k)}x(k)+E_{\bold{p}(k)}\varphi_{\bold{p}(k)}(F_{\bold{p}(k)}x(k))+B_{\bold{p}(k)}+D_{\bold{p}(k)}w(k)+R_{\bold{p}(k)}\varsigma(k),\\
y(k)=Cx(k),\end{array}\right.
\end{align}
where the additive noise $\varsigma(k)$ is a sequence of independent random vectors with multivariate standard normal distributions, and $\varphi_p:\R\rightarrow\R$ satisfies 
\begin{equation}\label{Eq_6a}
0\leq\frac{\varphi_p(c)-\varphi_p(d)}{c-d}\leq \bar a_p,~~~\forall c,d\in\R,c\neq d,
\end{equation}
for some $\bar a_p\in\R_{>0}\cup\{\infty\}$. 

We use the tuple
\begin{align}\notag
\Sigma=(A,B,C,D,E,F,R,\varphi),
\end{align}
to refer to the class of nonlinear switched systems of the form~\eqref{Eq_58a}, where $A = \{A_1,\dots, A_m\}, B = \{B_1,\dots, B_m\},\\ D= \{D_1,\dots, D_m\}, E = \{E_1,\dots, E_m\}, F= \{F_1,\dots, F_m\}, R = \{R_1,\dots, R_m\}, \varphi = \{\varphi_1,\dots, \varphi_m\}$, for the finite set of $P = \{1,\dots,m\}$.

\begin{remark}
	If $E_p$ is a zero matrix or $\varphi_p$ in~\eqref{Eq_58a} is linear including the zero function (i.e. $\varphi_p\equiv0$), one can remove or push the term $E_p\varphi_p(F_px)$ to $A_px$, and consequently the nonlinear tuple reduces to the linear one $\Sigma=(A,B,C,D,R)$. Then, every time we mention the tuple $\Sigma=(A,B,C,D,E,F,R,\varphi)$, it implicitly implies that $\varphi_p$ is nonlinear and $E_p$ is nonzero. 
\end{remark}

Here, we employ quadratic function of the form
\begin{align}\label{Eq_77a}
V_p(x,\hat x)=(x-\hat x)^T M_p(x-\hat x),\quad \forall p\in P,
\end{align}
where $M_p\succ0$ is a positive-definite matrix of an appropriate dimension. In order to show that a nominated $V$ employing $V_p$ in~\eqref{Eq_77a} is an SPSF from $\mathbb{\widehat G}(\widehat\Sigma)$ to $\mathbb{G}(\Sigma)$, we raise the following assumption on $\Sigma$. 

\begin{assumption}\label{As_31a}
	Assume that there exist constants $0<\bar \kappa_p <1$, $\pi_p \in \mathbb R_{>0}$, and matrix $ M_p\succ0$ such that the following inequality holds:
	\begin{align}\label{Eq_88a}
	&\begin{bmatrix}
	(1+2\pi_p)A_p^T M_pA_p& A_p^T M_pE_p\\
	E_p^TM_pA_p& (1+2\pi_p)E_p^T  M_pE_p
	\end{bmatrix}\preceq\begin{bmatrix}
	\bar\kappa_p M_p& -F_p^T\\
	-F_p & 2/\bar a_p
	\end{bmatrix}\!.
	\end{align}
\end{assumption}
\begin{remark}
	Note that for any \emph{linear} system $\Sigma=(A,B,C,D, R)$ with matrices $E_p$ and $F_p$ being identically zero, matrices $A_p$ being Hurwitz is sufficient to satisfy Assumption~\ref{As_31a}.
\end{remark}

Now, we provide another main result of this paper showing under which conditions a nominated $V$ using $V_p$ in~\eqref{Eq_77a} is an SPSF from $\mathbb{\widehat G}(\widehat\Sigma)$ to $\mathbb{G}(\Sigma)$.
\begin{theorem}\label{Thm_3a}
	Consider a global MDP $\mathbb{G}(\Sigma)$ associated with $\Sigma=(A,B,C,D,E,F,R,\varphi)$ and $\mathbb{\widehat G}(\widehat\Sigma)$ as its \emph{finite} abstraction with \emph{state} discretization parameter $\bar \delta$. Let $\epsilon>1$. If Assumption~\ref{As_31a} holds, and $\forall p\in P$, $~k_d \geq \epsilon\frac{\ln(\mu)}{\ln(1/\bar\kappa_p)}+1$, then 
	\begin{align}\label{function V: Nonlinear}
	V((x,p,l),(\hat x,p,l))=\frac{1}{{\bar\kappa_p}^{l/\epsilon}}V_p(x,\hat x),
	\end{align}
	with $V_p$ in~\eqref{Eq_77a} is an SPSF from $\mathbb{\widehat G}(\widehat\Sigma)$ to $\mathbb{G}(\Sigma)$.
\end{theorem}
The proof of Theorem~\ref{Thm_3a} is provided in the Appendix. Remark that $\mu$ employed in Theorem \ref{Thm_3a} is the one appearing in Assumption \ref{Assume: Switching}. Given the quadratic forms of $V_p$ in \eqref{Eq_77a}, $\forall p\in P$, we can always choose $\mu\geq1$ satisfying Assumption \ref{Assume: Switching} as discussed in Remark~\ref{existance mu}.

\section{Case Study}
In this section, to demonstrate the effectiveness of our proposed results, we first apply our approaches to a road traffic network in a circular cascade ring composed of $200$ identical cells, each of which has the length of $500$ meters with $1$ entry and $1$ way out, and construct compositionally a \emph{finite} MDP of the network. We employ the constructed finite abstraction as a substitute to compositionally synthesize policies keeping the density of traffic lower than 20 vehicles per cell. Finally, to show applicability of our results to switched systems accepting \emph{Multiple} Lyapunov functions with \emph{dwell-time}, we apply our proposed techniques to a \emph{fully interconnected} network of $500$ \emph{nonlinear} subsystems (totally $1000$ dimensions) and construct their \emph{finite} MDPs with guaranteed error bounds on their probabilistic output trajectories.

\subsection{Road Traffic Network}
\begin{figure}[ht]
	\begin{center}
		\includegraphics[width=5cm]{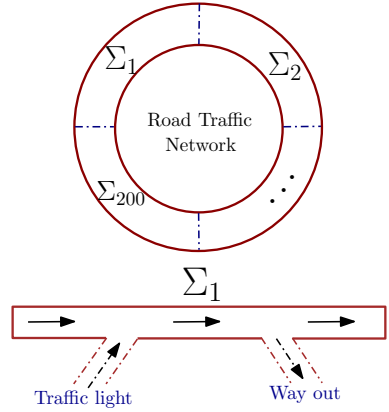}
		\caption{Model of a road traffic network in a circular cascade ring composed of $200$ identical cells, each of which has the length of $500$ meters with $1$ entry and $1$ way out.}
		\label{Fig2}
	\end{center}
\end{figure}

In this subsection, we apply our results to a road traffic network in a circular cascade ring which is composed of $200$ identical cells, each of which has the length of $500$ meters with $1$ entry and $1$ way out, as schematically depicted in Figure~\ref{Fig2}. The model of this case study is borrowed from~\cite{le2013mode} by including stochasticity in the model as an additive noise.

The entry is controlled by a traffic light,
that enables (green
light) or not (red light) the vehicles to pass. In this model the length of a cell is in kilometers
($0.5 ~km$), and the flow speed of the vehicles is $100$ kilometers per hour ($km/h$). Moreover,
during the sampling time interval $\tau=6.48$ seconds, it is assumed that $8$ vehicles pass the entry controlled by
the green light, 
and one quarter of vehicles goes out on the exit of each cell (ratio denoted by $q$). We want to observe the density of the traffic $x_i$, given in vehicles per cell, for each cell $i$ of the road. The set of modes is $P_i = \{1,2\},i\in\{1,\dots,n\}$ such that 
\begin{itemize}
	\item mode $1$ means traffic light is red;
	\item mode $2$ means traffic light is green.
\end{itemize}

Note that here we only have the traffic signals on the on-ramps. The dynamic of the interconnected system is described by:
\begin{align}\notag
\Sigma\!:\!\left\{\hspace{-1.5mm}\begin{array}{l}x(k+1) = A\,x(k) + B_{\bold{p}(k)} + \varsigma(k),\\
y(k)=x(k),\\
\end{array}\right.
\end{align}
where $A$ is a matrix with diagonal elements $ a_{ii}=(1-\frac{\tau \nu_i}{L_i}-q)$, $i\in\{1,\ldots,n\}$, off-diagonal elements $ a_{i+1,i}=\frac{\tau \nu_i}{L_i}$, $i\in \{1,\ldots,n-1\}$, $a_{1,n} = \frac{\tau \nu_n}{L_n}$, and all other elements are identically zero. Moreover, $B_p=[b_{1p_1};\ldots;b_{np_n}]$, $x(k)=[x_1(k);\ldots;x_n(k)]$, $\varsigma(k)=[\varsigma_1(k);\ldots;\varsigma_n(k)]$, and
\begin{align}\notag
b_{ip_i} =\left\{\hspace{-1.5mm}\begin{array}{l} 0,\quad\quad\quad \text{if}~~~~  p_i = 1,\\
8,\quad\quad \quad\text{if}~~~~  p_i = 2.\\
\end{array}\right.
\end{align}
Furthermore, the additive noise $\varsigma(k)$ is a sequence of independent random vectors with multivariate standard normal distributions (i.e., mean zero and covariance matrix identity). Now, by introducing the individual cells $\Sigma_i$ described as
\begin{align}\notag
\Sigma_i:\left\{\hspace{-1.5mm}\begin{array}{l}x_i(k+1) = (1-\frac{\tau \nu_i}{L_i}-q)\,x_i(k) + D_iw_i(k) +b_{i{\bold{p}_i(k)}}+\varsigma_i(k),\\
y_i(k)=x_i(k),\\
\end{array}\right.
\end{align}
where $D_i = \frac{\tau \nu_{i-1}}{L_{i-1}}$ (with $\nu_0 = \nu_n$, $L_0 = L_n$) and $w_i(k) = y_{i-1}(k)$ (with $y_0 = y_n$), one can readily verify that $\Sigma=\mathcal{I}(\Sigma_1,\ldots,\Sigma_N)$, equivalently $\Sigma=\mathcal{I}(\mathbb{G}(\Sigma_1),\ldots,,\mathbb{G}(\Sigma_N))$. Note that we consider sets $X_i = W_i = [0~~20]$, $\forall i \in \{1,\dots, n\}$. Since the dynamic of the system is linear, condition~\eqref{Eq_88a} reduces to, 
\begin{align}\label{Eq9a}
(1+2\pi_{i})A_{i}^T M_iA_{i}\preceq  \bar\kappa_i  M_i, 
\end{align}
which is nothing more than stability of each cell $i$. Note that in this example $V_p = V_{p'}, \forall p,p' \in P$  (i.e., common Lyapunov function). Then one can readily verify that this condition is satisfied with $M_i=1$, $\pi_i = 0.85$, $\bar\kappa_i = 0.41$ $\forall i\in\{1,\ldots,n\}$, and the function $V_i(x_i,\hat x_i)=(x_i-\hat x_i)^2$ is an SPSF from $\mathbb{\widehat G}(\widehat\Sigma_i)$ to $\mathbb{G}(\Sigma_i)$ satisfying condition \eqref{Eq_2a} with $\alpha_{i}(s)=s^2$ and condition \eqref{Eq_3a} with $\kappa_i=0.99$, $\rho_{\mathrm{int}i}(s)=0.72s^2$, $\forall s\in \mathbb R_{\ge0}$, and $\psi_i = 84.96\,\bar\delta_i^2$.

Now we check the small-gain condition~\eqref{Assump: Kappa1} that is required for the compositionality result.
By taking $\sigma_i(s) = s$, $\forall i\in\{1,\ldots,n\}$, condition~\eqref{Assump: Kappa1} and as a result condition \eqref{compositionality1} are always satisfied without any restriction on the number of cells.
Hence, $V(x,\hat x)=\max_{i} (x_i-\hat x_i)^2$ is an SSF from  $\mathbb{\widehat G}(\widehat\Sigma)$ to $\mathbb{G}(\Sigma)$ satisfying conditions \eqref{eq:lowerbound2} and \eqref{eq6666}  with $\alpha(s)=s^2$, $\kappa=0.99$, and $\psi = 84.96 \,\bar\delta^2$.

We take the state and internal input discretization parameters as $0.02$. Hence, we have $n_{x_i} = n_{w_i} = 1000$. By taking the initial states of the interconnected systems $\Sigma$ and $ \widehat \Sigma$ as $10\mathds{1}_{200}$, we guarantee that the distance between trajectories of $\Sigma$ and of $\widehat \Sigma$ will not exceed $\varepsilon = 1$ during the time horizon $T_d=15$ with the probability at least $88\%$, i.e.,
\begin{align}\notag
\mathds{P}(\Vert y_{a\hat\nu}(k)-\hat y_{\hat a \hat\nu}(k)\Vert\le 1,\,\, \forall k\in[0,15])\ge 0.88.
\end{align}

Note that for the construction of finite abstractions, we have selected the center of partition sets as representative points. We do not need any constraint on the shape of the
partition sets in general in constructing finite MDPs.
For the sake of an easy implementation, the partition sets are considered hyper-intervals and the center of them as
their representative points.
Moreover, we assume $\hat Y_{ij}= \hat W_{ji}$, i.e., the overall error in~\eqref{overall-error} reduces to $\psi:=\max_{i}\sigma_i^{-1}(\psi_i)$.

\subsection{Compositional Controller Synthesis}

\begin{figure}
	\centering
	\includegraphics[width=8.4cm]{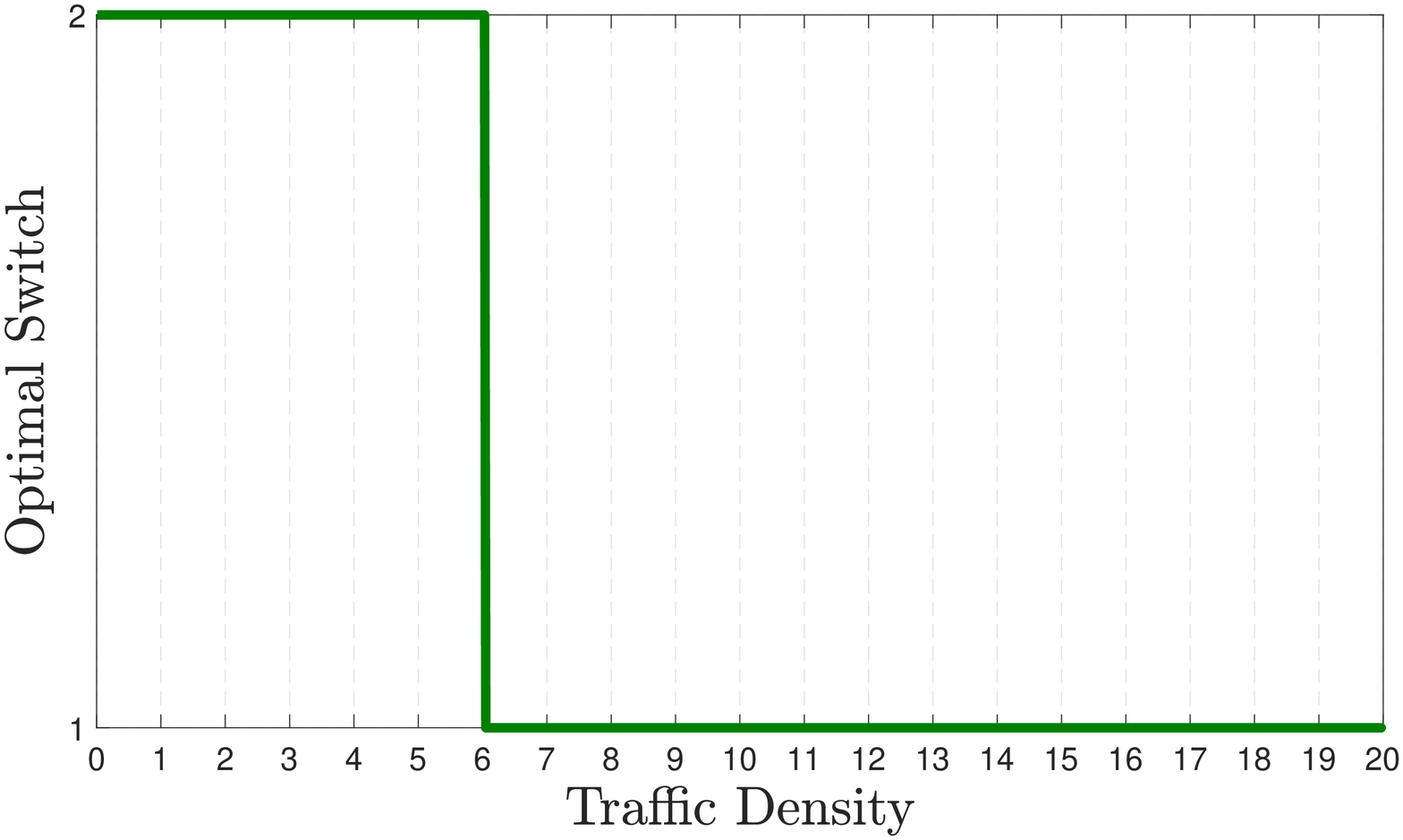}
	\includegraphics[width=8.4cm]{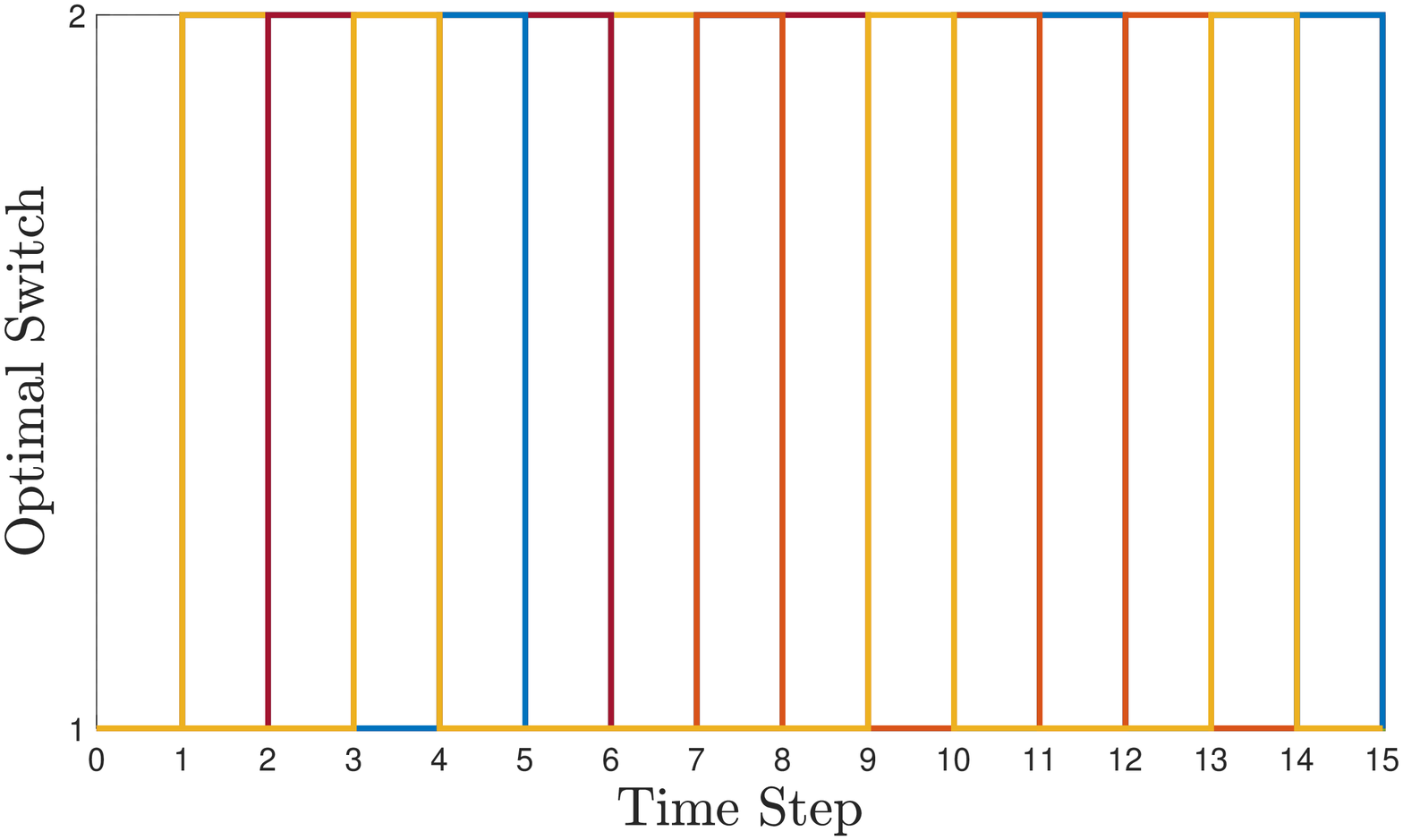}
	\includegraphics[width=8.4cm]{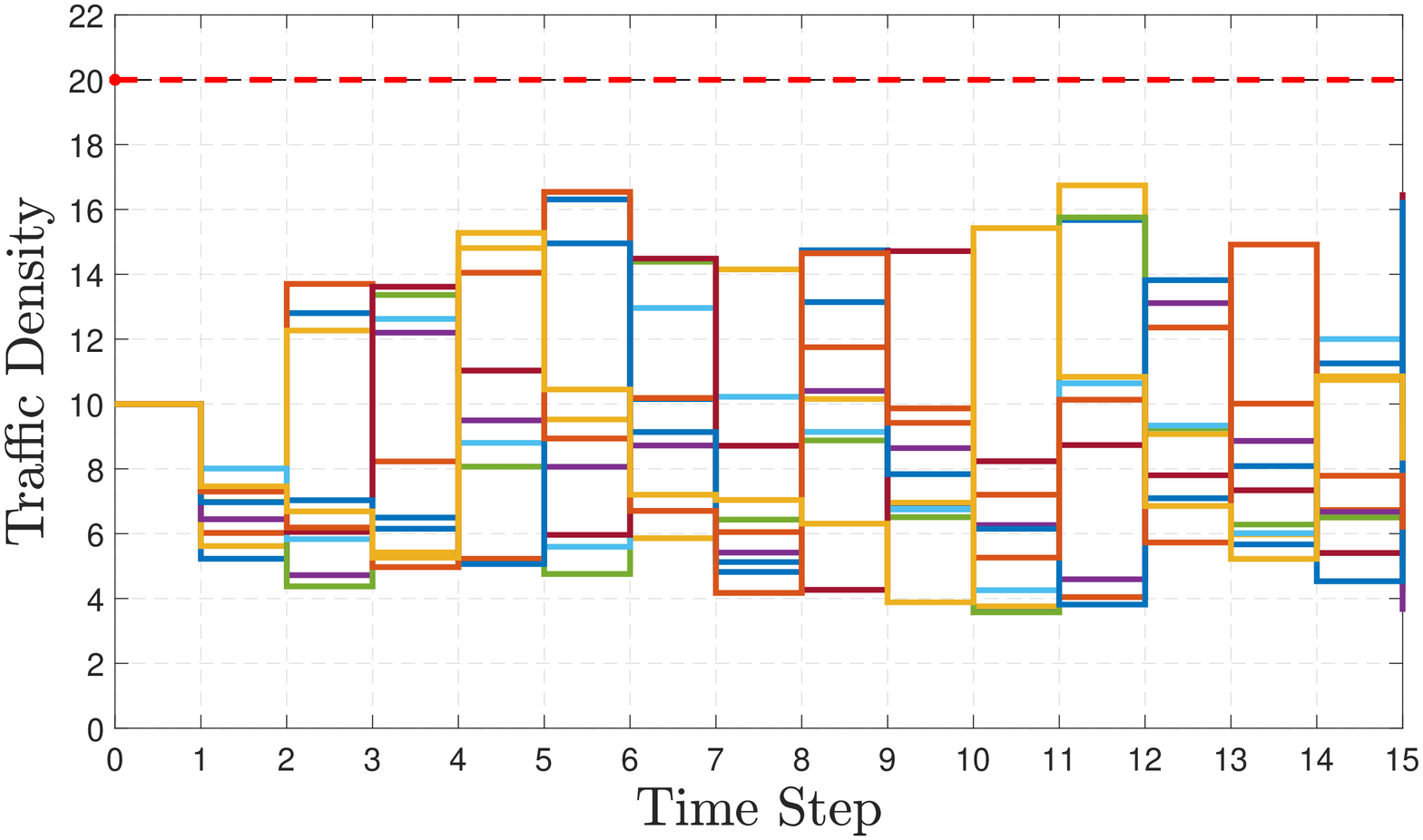}
	\caption{Top: Optimal switch for a representative cell in a network of $200$ cells. Middle: Optimal switch w.r.t. time for a representative cell with $10$ different noise realizations. Bottom: Closed-loop state trajectories of a representative cell with $10$ different noise realizations.}
	\label{Optimal_Policy}
\end{figure}

Let us now synthesize a controller for $\Sigma$ via the abstraction $\mathbb{\widehat G}(\widehat\Sigma)$ such that the \emph{safety} controller maintains the density of traffic lower than $20$ vehicles per cell. The idea here is to first design a local controller for the abstraction $\mathbb{\widehat G}(\widehat\Sigma_i)$, and then refine it back to system $\Sigma_i$. Consequently, a controller for the interconnected system  $\Sigma$ would be a vector such that each of its components is the controller for systems $\Sigma_i$. We employ here software tool \software \cite{FAUST15} by doing some modification to accept internal inputs as disturbances, and synthesize a controller for $\Sigma$ by choosing the standard deviation of the noise $\sigma_i = 0.83$, $\forall i\in\{1,\ldots,n\}$.
Optimal switch for a representative cell in a network of $N = 200$ cells is plotted in Figure~\ref{Optimal_Policy} top. Optimal switch here is sub-optimal for each subsystem and is obtained by assuming that other subsystems do not violate the safety specification. Optimal switch w.r.t. time for a representative cell with different noise realizations is also illustrated in Figure~\ref{Optimal_Policy} middle, with $10$ realizations. Moreover, closed-loop state trajectories of the representative cell with different noise realizations are illustrated in Figure~\ref{Optimal_Policy} bottom. 

\subsection{Memory Usage and Computation Time}

Now we discuss the memory usage and computation time of constructing finite MDPs in both monolithic and compositional manners. The monolithic finite MDP would be a matrix with the dimension of $(n_{x_i}^N\times 2^N)\times n_{x_i}^N$ with $n_{x_i} = 1000$ and $N=200$. By allocating $8$ bytes for each entry of the matrix to be stored as a double-precision floating point, one needs a memory of $\frac{8\times1000^{200}\times2^{200}\times 1000^{200}}{10^9} \approx 10^{1252}$ GB for building the finite MDP in the monolithic manner which is impossible in practice. Now we proceed with the compositional construction of finite MDPs proposed in this work. The constructed MDP for each subsystem here is a matrix with the dimension of $(n_{x_i}\times 2\times n_{w_i})\times n_{x_i}$ with $n_{x_i} = n_{w_i} = 1000$. This has the memory usage of $\frac{8\times1000\times 2\times1000\times1000}{10^9} =16$ GB.
We can compute such a finite MDP with the software tool \software, which takes $112$ seconds on a machine with Windows operating system (Intel i7@3.6GHz CPU and 16 GB of RAM).

A comparison on the required memory for the construction of finite MDPs between the monolithic and compositional manners for different state discretization parameters is provided in Table~\ref{Table}.
As seen, in order to provide even a very weak closeness guarantee of $2\%$ between trajectories of $\Sigma$ and of $\widehat \Sigma$, the required memory in the monolithic fashion is $10^{972}$ GB which is still impossible in practice. This implementation clearly shows that the proposed compositional approach in this work significantly mitigates the curse of dimensionality problem in constructing finite MDPs monolithically. In particular, in order to quantify the probabilistic closeness between two networks $\Sigma$ and $\widehat\Sigma$ via inequality~\eqref{Eq_25} as provided in Table~\ref{Table}, one needs to only build finite MDPs of individual subsystems (i.e., $\widehat\Sigma_i$), construct an SPSF between each $\Sigma_i$ and $\widehat\Sigma_i$, and then employ the proposed compositionality results of the paper to build an SSF between $\Sigma$ and $\widehat\Sigma$.

\begin{table}[ht]
	\caption{Required memory for the construction of finite MDPs in both monolithic and compositional manners for different state discretization parameters.}
	\centering
	\begin{tabular}{c c c c}
		\hline
		$\bar\delta$ & Closeness &  $\widehat\Sigma_i$ (GB)& $\widehat\Sigma$ (GB)\\ [0.5ex]
		\hline
		$0.01$ & $97\%$ & $128$ & $10^{1372}$ \\
		$0.02$ & $88\%$ &$16$ & $10^{1252}$ \\
		$0.03$ & $75\%$ & $4.72$ & $10^{1181}$\\
		$0.04$ & $60\%$ & $2$ & $10^{1131}$ \\
		$0.05$ & $44\%$ & $1.02$ & $10^{1092}$ \\
		$0.06$ & $30\%$ & $0.59$ & $10^{1061}$ \\ 
		$0.07$ & $19\%$ & $0.37$ & $10^{1033}$ \\
		$0.08$ & $11\%$ & $0.25$ & $10^{1011}$ \\
		$0.09$ & $5\%$ & $0.17$ & $10^{990}$ \\
		$0.1$ & $2\%$ & $0.12$ & $10^{972}$ \\[1ex]
		\hline
	\end{tabular}
	\label{Table}
\end{table}

\subsection{Comparisons with DBN Approach of \cite{SAM17}}

\begin{figure}
	\centering
	\includegraphics[scale=0.26]{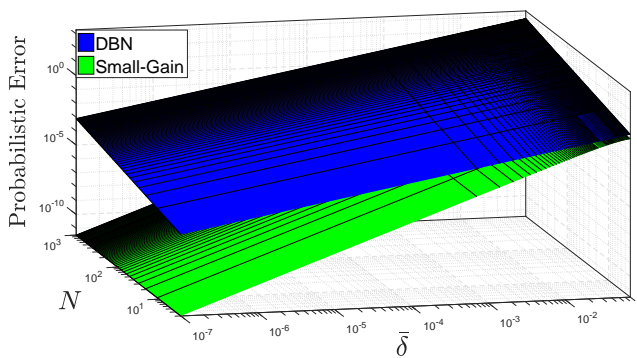}
	\caption{Comparison of the probabilistic error bound in \eqref{Eq_25} provided by our approach based on $\max$ small-gain conditions with that of \cite{SAM17} based on DBN. Plots are in the logarithmic scale for a fixed $\varepsilon = 1$, $\sigma_i = 0.83$, and  $T_d = 15$.}
	\label{DBN_delta_N}
\end{figure}

\begin{figure}
	\centering
	\includegraphics[scale=0.26]{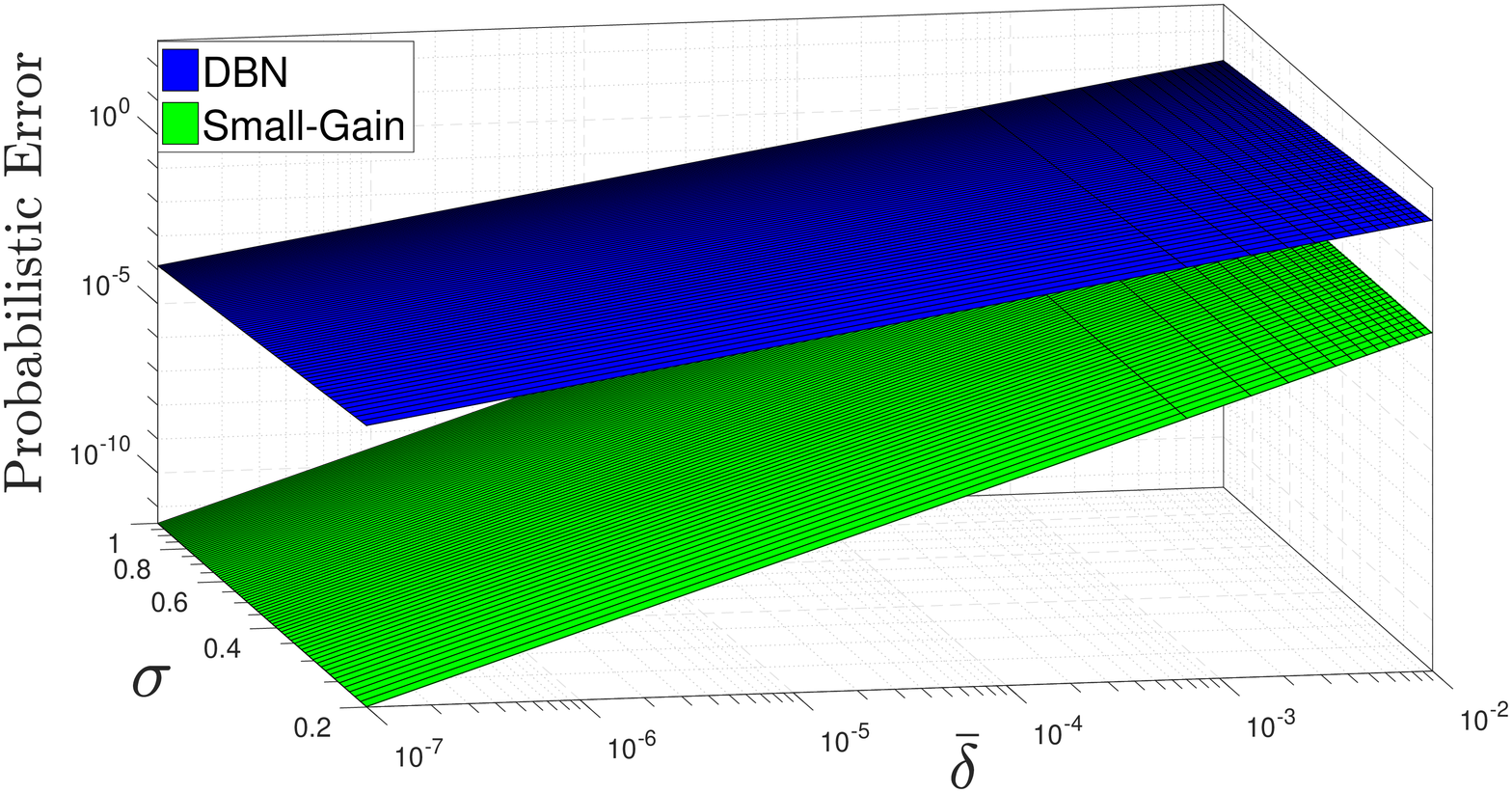}
	\caption{Comparison of the probabilistic error bound in \eqref{Eq_25} provided by our approach based on $\max$ small-gain conditions with that of \cite{SAM17} based on DBN. Plots are in the logarithmic scale for a fixed $N = 200$, $\varepsilon = 1$, and $T_d = 15$.}
	\label{DBN_delta_sigma}
\end{figure}

\begin{figure}
	\centering
	\includegraphics[scale=0.26]{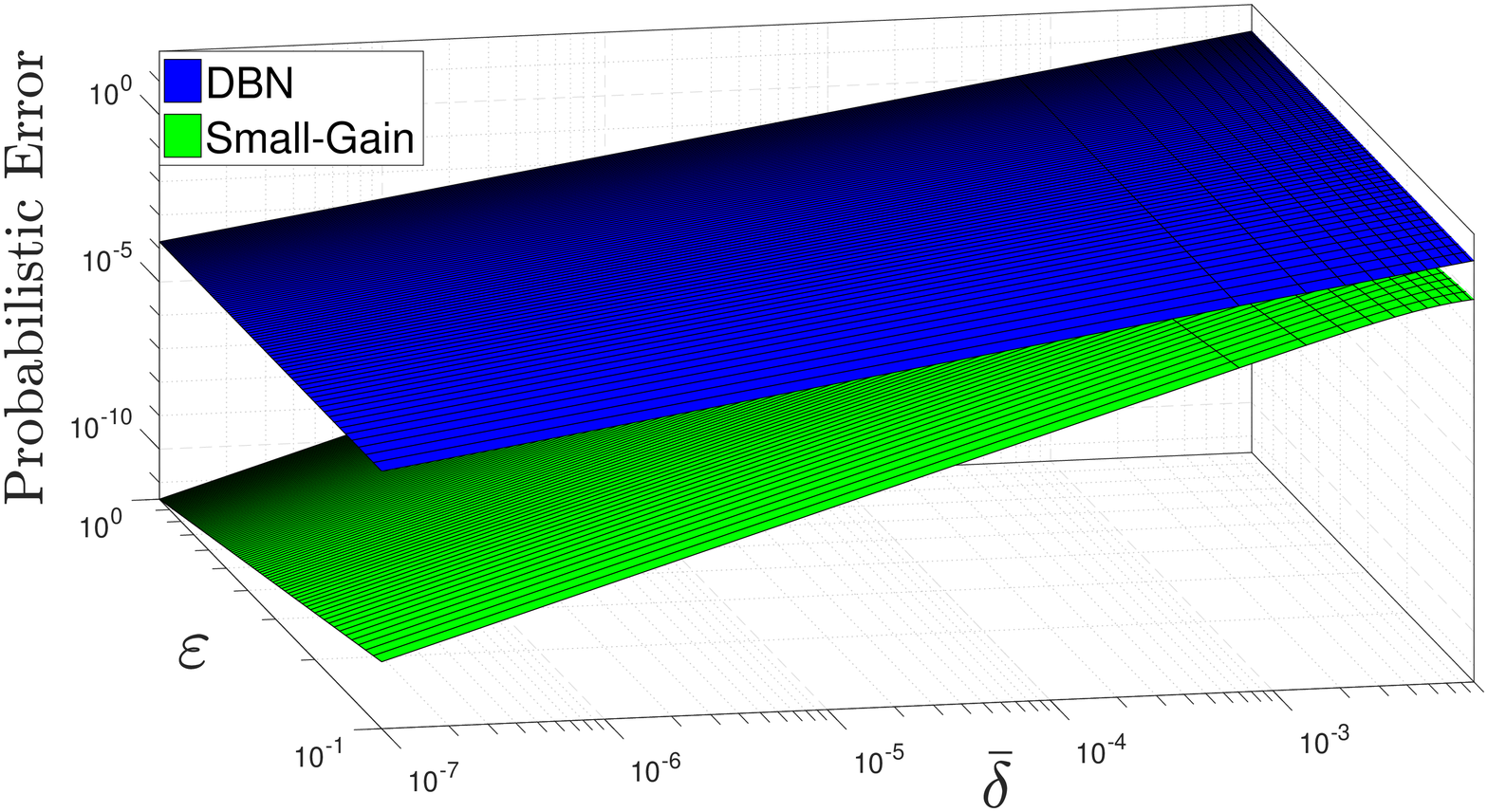}
	\caption{Comparison of the probabilistic error bound in \eqref{Eq_25} provided by our approach based on $\max$ small-gain conditions with that of \cite{SAM17} based on DBN. Plots are in the logarithmic scale for a fixed $N = 200$, $\sigma_i = 0.83$, and $T_d = 15$.}
	\label{DBN_delta_varepsilon}
\end{figure}

We first compare the probabilistic closeness guarantees provided by our approach with that of~\cite{SAM17}. Note that our results are based on $\max$ small-gain conditions while~\cite{SAM17} employs dynamic Bayesian network (DBN) to capture the dependencies between subsystems.
The comparison is shown in Figures~\ref{DBN_delta_N}-\ref{DBN_delta_varepsilon} in the logarithmic scale.
In Figure~\ref{DBN_delta_N}, we have fixed the confidence bound $\varepsilon = 1$, the standard deviation of the noise $\sigma_i = 0.83$, the time horizon $T_d = 15$, and plotted the error as a function of  the state discretization parameter $\bar\delta$ and the number of subsystems $N$. As seen, by increasing the number of subsystems, our error provided in~\eqref{Eq_25} does not change since the overall $\psi$ is independent of the size of the network (i.e. $N$), and is computed only based on the maximum $\psi_i$ of subsystems instead of being a linear combination of them which is the case in~\cite{SAM17}.
In Figure~\ref{DBN_delta_sigma}, we have fixed $N = 200,$ $\varepsilon = 1$, $T_d = 15$, and plotted the error as a function of $\bar\delta$ and $\sigma$. Our error in \eqref{Eq_25} is independent of $\sigma$ while the error in~\cite{SAM17} grows when $\sigma$ goes to zero.
In Figure~\ref{DBN_delta_varepsilon}, we have fixed $N = 200,$ $\sigma_i= 0.83$, $T_d = 15$, and plotted the error as a function of $\bar\delta$ and $\varepsilon$. The error in \cite{SAM17} is independent of $\varepsilon$ while our error increases when $\varepsilon$ goes to zero.

In conclusion, the proposed approach in~\cite{SAM17} is more general than our setting here. It does not require original systems to be incremental input-to-state stable ($\delta$-ISS) and only the Lipschitz continuity of the associated stochastic kernels is enough for validity of the results. The refinement does not require running the abstract systems and obtaining the input according to an interface function.
On the other hand, the abstraction error in~\cite{SAM17} depends on the number of subsystems and also the Lipschitz constants of the stochastic kernels associated with the system. Thus, our approach outperforms the results in~\cite{SAM17} for large-scale stochastic systems with small standard deviation of the noise as long as the imposed assumptions are satisfied.

\subsection{Comparisons with Dissipativity Approach in \cite{lavaei2017HSCC}}
\begin{figure}
	\centering
	\includegraphics[scale=0.26]{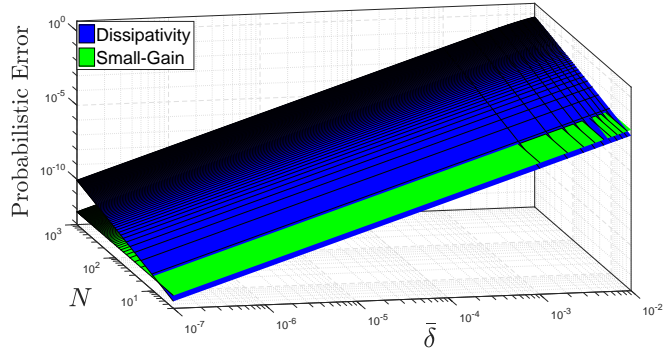}
	\caption{Comparison of the probabilistic error bound in \eqref{Eq_25} provided by our approach based on $\max$ small-gain conditions with that of \cite{lavaei2017HSCC} based on dissipativity-type reasoning. Plots are in the logarithmic scale for a fixed $\varepsilon = 1$, and  $T_d = 15$.}
	\label{Dissipativity_delta_N}
\end{figure}

Since the presented road traffic network admits a common Lyapunov function, our results recover the ones proposed in \cite{lavaei2017HSCC} by considering switching signals as discrete inputs.
The comparison is shown in Figure~\ref{Dissipativity_delta_N} in the logarithmic scale. We have fixed $\varepsilon = 1$, $T_d = 15$, and plotted the error as a function of $\bar\delta$ and the number of subsystems $N$. By increasing the number of subsystems, the error in~\eqref{Eq_25} does not change since the overall $\psi$ is independent of $N$, and is computed only based on the maximum of $\psi_i$ of subsystems instead of being a linear combination of them which is the case in~\cite{lavaei2017HSCC}. Nevertheless, for networks with small number of subsystems, the proposed errors in~\cite{lavaei2017HSCC} are slightly better than the ones provided in this work. This issue is expected and the reason is due to the conservatism nature of the approach that we employ here (\cite[Theorem 1]{Abdallah2017compositional}) to transfer the additive form of our pseudo-simulation functions to a $\max$ form (cf.~\eqref{max form}),
but with the gain of providing an overall error for the network only based on the maximum error of subsystems instead of a linear combination of them. Thus, our proposed results here outperform the ones in~\cite{lavaei2017HSCC} for large-scale stochastic switched systems admitting a common Lyapunov function.


\subsection{Switched Systems Accepting Multiple Lyapunov Functions with Dwell-Time}
\begin{figure}
	\centering
	\includegraphics[width=4.7cm]{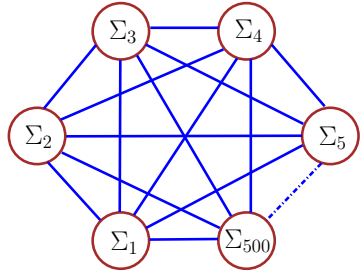}
	\caption{A \emph{fully interconnected} network of $500$ \emph{nonlinear} components (totally $1000$ dimensions).}
	\label{Case_Study}
\end{figure}

In order to show applicability of our results to switched systems accepting \emph{multiple} Lyapunov functions with a \emph{dwell-time}, we apply our proposed techniques to a \emph{fully interconnected} network of $500$ \emph{nonlinear} subsystems in the form of~\eqref{Eq_58a} (totally $1000$ dimensions), as illustrated in Figure~\ref{Case_Study}. The model of the system does not have a common Lyapunov function because it exhibits unstable behaviors for different switching signals~\cite{liberzon2003switching} (i.e., if one periodically switches between different modes, the trajectory goes to infinity). The dynamic of the interconnected system is described by:
\begin{equation*}
\Sigma:\left\{\hspace{-1.5mm}\begin{array}{l}{x}(k+1)= A_{\bold{p}(k)}x(k)+B_{\bold{p}(k)}+\varphi(x(k))+R\varsigma(k),\\
y(k)=x(k),\end{array}\right.
\end{equation*}
where 
\begin{align}\notag
A_{\bold{p}(k)}&=\begin{bmatrix}\bar A_{{pi}} & \tilde A  & \cdots & \cdots & \tilde A  \\  \tilde A  & \bar A_{{pi}} & \tilde A  & \cdots & \tilde A  \\ \tilde A  & \tilde A  & \bar A_{{pi}} & \cdots & \tilde A  \\ \vdots &  & \ddots & \ddots & \vdots \\ \tilde A  & \cdots & \cdots & \tilde A  & \bar A_{{pi}}\end{bmatrix}_{n\times n}\!\!\!\!\!\!\!\!\!\!\!\!,\\\notag
\tilde A&= \begin{bmatrix}0.015 & 0 \\  0 & 0.015\end{bmatrix}\!\!,
\quad \bar A_{{p_i}} =\left\{\hspace{-1.7mm}\begin{array}{l} \begin{bmatrix}0.05 & 0 \\  0.9 & 0.03\end{bmatrix}\!,\quad \vspace{1mm}~\text{if}~~  p_i = 1,\\
\begin{bmatrix}0.02 & -1.2 \\  0 & 0.05\end{bmatrix}\!,\quad\text{if}~~  p_i = 2.\\
\end{array}\right.
\end{align}

Moreover, we choose $R = \mathsf{diag}(\mathds{1}_{2},\ldots,\mathds{1}_{2})$, $\varphi(x)=[0.1\mathds{1}_{2}\varphi_1(0.1\mathds{1}_{2}^Tx_1(k));\ldots;0.1\mathds{1}_{2}\varphi_N(0.1\mathds{1}_{2}^Tx_N(k))]$, and $\varphi_i(x) = sin(x)$, $\forall i\in\{1,\ldots,N\}$. Note that functions $\varphi_i$ satisfy condition \eqref{Eq_6a} with $\bar a_{{p_i}} = 1$. We fix here $N=500$. Furthermore, $B_p=[b_{1p_1};\ldots;b_{Np_N}]$ such that
\begin{align}\notag
b_{ip_i} =\left\{\hspace{-1.7mm}\begin{array}{l} \begin{bmatrix}-0.9 \\  0.5 \end{bmatrix}\!,\quad\quad \quad\vspace{1mm}\text{if}~~~~  p_i = 1,\\
\begin{bmatrix}0.9 \\  -0.2 \end{bmatrix}\!,\quad\quad\quad \text{if}~~~~  p_i = 2.\\
\end{array}\right.
\end{align}
We partition $x(k)$ as $x(k)=[x_1(k);\ldots;x_N(k)]$ and $\varsigma(k)$ as $\varsigma(k)=[\varsigma_1(k);\ldots;\varsigma_N(k)]$, where $x_i(k),\varsigma_i(k)\in\R^{2}$.
Now, by introducing the individual subsystems $\Sigma_i$ described as
\begin{align}\notag
\Sigma_i:\left\{\hspace{-1.5mm}\begin{array}{l}x_i(k+1)= \bar A_{{\bold{p}_i(k)}}x_i(k) +b_{i{\bold{p}_i(k)}}+D_iw_i(k)+ 0.1\mathds{1}_{2}\varphi_i(0.1\mathds{1}_{2}^Tx_i(k))+\mathds{1}_{2}\varsigma_i(k),\\
y_i(k)=x_i(k),\\
\end{array}\right.
\end{align}
where
\begin{align}\notag 
D_i = [\tilde A;\dots;\tilde A]^T_{2\times (n-2)},\quad w_i(k)=[{y_{i1};\ldots;y_{i(i-1)};y_{i(i+1)};\ldots;y_{iN}}], ~i \in \{1,\ldots,N\}, 
\end{align}
one can readily verify that $\Sigma=\mathcal{I}(\Sigma_1,\ldots,\Sigma_N)$, equivalently $\Sigma=\mathcal{I}(\mathbb{G}(\Sigma_1),\ldots,\mathbb{G}(\Sigma_N))$. One can also verify that, $\forall i\in\{1,\ldots,N\}$, condition~\eqref{Eq_88a} is satisfied with
\begin{align}\notag
&\text{for} \,p_i = 1\!: \,M_{{p_i}} = \begin{bmatrix}1.311 & 0.001 \\  0.001 & 0.492\end{bmatrix}\!, \bar\kappa_{{p_i}} = 0.7, \pi_{{p_i}} = 0.5,\\\notag
&\text{for} \,p_i = 2\!: \,M_{{p_i}} = \begin{bmatrix}0.4 & 0.01 \\  0.01 & 1.49\end{bmatrix}\!, \bar\kappa_{{p_i}} = 0.7, \pi_{{p_i}} = 0.4.
\end{align}By taking $\epsilon =1.75$ and choosing $\mu = 3.27$, one can get the dwell-time $k_d = 7$. Hence, $V_i((x_i,p_i,l_i),(\hat x_i,p_i,l_i))=\frac{1}{\bar\kappa_{p_i} ^ {l/1.75}}(x_i-\hat x_i)^TM_{ip_i}(x_i-\hat x_i)$ is an SPSF from $\mathbb{\widehat G}(\widehat\Sigma_i)$ to $\mathbb{G}(\Sigma_i)$ satisfying condition \eqref{Eq_2a} with  $\alpha_{i}(s)=0.2s^2$ and condition \eqref{Eq_3a} with $\kappa_i=0.99$, $\rho_{\mathrm{int}i}(s)=0.19s^2$, $\forall s\in \mathbb R_{\ge0}$, and $\psi_i = 2266\,\bar\delta_i^2$.

Now we the check small-gain condition~\eqref{Assump: Kappa1} that is required for the compositionality result.
By taking $\sigma_i(s) = s$, $\forall i\in\{1,\ldots,N\}$, condition~\eqref{Assump: Kappa1} and as a result condition \eqref{compositionality1} are satisfied. Hence, $V((x,p,l),(\hat x,p,l))=\max_{i} \{\frac{1}{\bar\kappa_{p_i} ^ {l/1.75}}(x_i-\hat x_i)^TM_{ip_i}(x_i-\hat x_i)\}$  is an SSF from $\mathbb{\widehat G}(\widehat\Sigma)$ to $\mathbb{G}(\Sigma)$ satisfying conditions \eqref{eq:lowerbound2} and \eqref{eq6666}  with $\alpha(s)=0.2s^2$, $\kappa=0.99$, and $\psi = 2266\,\bar\delta^2$.

By taking the state set discretization parameter $\bar\delta_i = 0.001$, and taking the initial states of the interconnected systems $\Sigma$ and $ \widehat \Sigma$ as $\mathds{1}_{1000}$, we guarantee that the distance between trajectories of $\Sigma$ and of $\widehat \Sigma$ will not exceed $\varepsilon = 1$ during the time horizon $T_d=10$ with the probability at least $90\%$, i.e.,
\begin{align}\notag
\mathds{P}(\Vert y_{a\hat\nu}(k)-\hat y_{\hat a \hat\nu}(k)\Vert\le 1,\,\, \forall k\in[0,10])\ge 0.9.
\end{align}
\subsection{Analysis on Probabilistic Closeness Guarantee}
\begin{figure}[ht]
	\centering
	\includegraphics[scale=0.26]{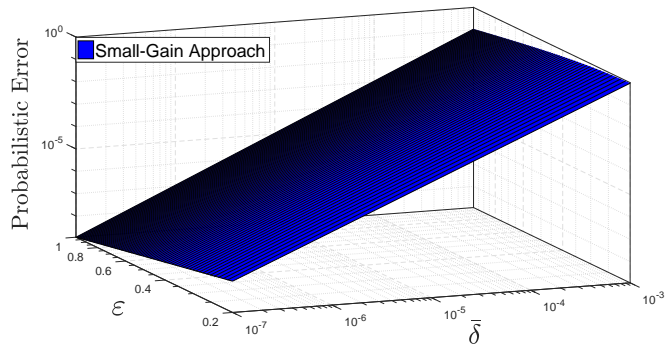}\vspace{-0.4cm}
	\caption{Probabilistic error bound proposed in \eqref{Eq_25} based on $\bar\delta$ and $\varepsilon$. Plot is in the logarithmic scale for $T_d = 10$. The probabilistic closeness guarantee is improved by either decreasing the state discretization parameter $\bar\delta$ or increasing the confidence bound $\varepsilon$.}
	\label{Fig6}
\end{figure}

In order to have a practical analysis of the probabilistic closeness guarantee, we plotted  in Figure \ref{Fig6} the probabilistic error bound provided in~\eqref{Eq_25} in terms of the state discretization parameter $\bar\delta$ and the confidence bound $\varepsilon$. As seen, the probabilistic closeness guarantee is improved by either decreasing $\bar\delta$ or increasing $\varepsilon$. Note that the constant $\psi$ in~\eqref{Eq_25} is formulated based on the state discretization parameter $\bar\delta$ as in~\eqref{Eq_305a}.
It is worth mentioning that there are some other parameters in~\eqref{Eq_25} such as $\mathcal{K}_\infty$ function $\alpha$, and the value of SSF $V$ at initial conditions $a, \hat a, p_0, l_0$ which can also improve the proposed bound for given values of $T_d$ and initial conditions of the system.

\section{Discussion}
In this paper, we provided a compositional approach for the construction of finite MDPs for  networks of discrete-time stochastic switched systems.  First, we introduced new notions of stochastic pseudo-simulation and simulation functions in order to quantify the probabilistic distance between concrete stochastic switched subsystems and their finite abstractions and their interconnections, respectively. Then we leveraged sufficient small-gain type conditions for the compositional quantification of the probabilistic distance between the interconnection of stochastic switched subsystems and that of their finite abstractions. Furthermore, we showed that under an incremental input-to-state stability property, one can construct finite MDPs of the concrete models for the \emph{general setting} of nonlinear stochastic switched systems. We also proposed an approach to construct finite MDPs together with their corresponding stochastic pseudo-simulation functions for a particular class of discrete-time \emph{nonlinear} stochastic switched systems. Finally, we applied our approaches to a road traffic network in a circular cascade ring composed of $200$ cells, and constructed compositionally a \emph{finite} MDP of the network. We employed the constructed finite abstraction as a substitute to compositionally synthesize policies keeping the density of the traffic lower than $20$ vehicles per cell. We also applied our proposed techniques to a \emph{fully interconnected} network of $500$ \emph{nonlinear} subsystems (totally $1000$ dimensions) accepting \emph{multiple} Lyapunov functions with the \emph{dwell-time}, and constructed their \emph{finite} MDPs with guaranteed error bounds. We benchmarked our proposed results against the ones available in the literature.

\bibliographystyle{alpha}
\bibliography{biblio}

\section{Appendix}

\begin{IEEEproof}\textbf{(Proposition~\ref{Proposition})}
In order to show that global MDP $\mathbb{G}(\Sigma)$ in Definition~\ref{def: gMDP} is itself an MDP, we need to elaborate on this issue that $\mathbb{X}$ is itself a Borel space. Since $X$ defined in~\eqref{eq:dt-SS} is a Borel space, one can readily verify that its Cartesian product by other discrete spaces as $\mathbb{X} = X \times P \times \{0,\dots,k_d-1\}$  is also a Borel space~\cite{APLS08}. Then the global MDP $\mathbb{G}(\Sigma) = (\mathbb{X},\mathbb{U},\mathbb{W},\varsigma,\mathbb{F},\mathbb{Y},\mathbb{H})$ can be \emph{equivalently} represented as an MDP
\begin{equation}\notag
\mathbb{G}(\Sigma) =(\mathbb{X},\mathbb{U},\mathbb{W},\mathbb{T_{\mathsf x}},\mathbb{Y},\mathbb{H}),	
\end{equation}
where the map $\mathbb{T_{\mathsf x}}:\mathcal B(\mathbb{X})\times \mathbb{X}\times \mathbb{U}\times \mathbb{W}\rightarrow[0,1]$,
is a conditional stochastic kernel that assigns to any $x \in \mathbb{X}$, $\nu\in \mathbb{U}$, and $w\in \mathbb{W}$ a probability measure $\mathbb{T_{\mathsf x}}(\cdot | x,\nu, w)$
on the measurable space
$(\mathbb{X},\mathcal B(\mathbb{X}))$
so that for any set $\mathcal{A} \in \mathcal B(\mathbb{X})$,
$$\mathds{P} (x(k+1)\in \mathcal{A}|x(k),\nu(k),w(k)) = \int_\mathcal{A} \mathbb{T_{\mathsf x}} (d x'|x(k),\nu(k),w(k)).$$
Moreover,
\begin{align}\notag
(p',l'):=
\begin{cases}
(p,l+1) \!\!\quad\quad\quad\text{if}\quad l<k_d-1,\\
(p,k_d-1) \quad\quad\text{if}\quad l=k_d-1,\\
(\neq p,0) ~\quad\quad\quad\!\text{if}\quad l=k_d-1,\\
\end{cases}
\end{align}
or equivalently,
\begin{align}\notag
\nu:=
\begin{cases}
\text{no switch} \!\!\!\!\!\!\!\quad\quad\quad\quad\quad\quad\text{if}\quad l<k_d-1,\\
\{1,2,\dots,m\} \!\!\!\!\!\!\!\!\quad\quad\quad\quad\quad\!\text{if}\quad l=k_d-1.\\
\end{cases}
\end{align}
Then the global MDP $\mathbb{G}(\Sigma)$ in Definition~\ref{def: gMDP} is itself an MDP.
Now we elaborate on the fact that the output trajectories of $\Sigma$ defined in~\eqref{Eq_1a} and of $\mathbb{G}(\Sigma)$ are equivalent. Given an initial state $x_0$, a switching signal $\bold{p}:\mathbb N \rightarrow P$, an internal input $w(\cdot)$, and a realization of the noise $\varsigma(\cdot)$, one can uniquely map the output trajectory of $\Sigma$ to an output trajectory of $\mathbb{G}(\Sigma)$. Moreover, if we pick $p_0\in P$ as the initial mode of the system and $l_0 = 0$, the output trajectory of $\mathbb{G}(\Sigma)$ can be uniquely projected to an output trajectory of $\Sigma$. Then one can uniquely map the output trajectory of $\Sigma$ to an output trajectory of $\mathbb{G}(\Sigma)$ and vice versa, for the same initial conditions.
	
\end{IEEEproof}	

\begin{IEEEproof}\textbf{(Theorem~\ref{Thm_1a})}
	For any $(x,p,l)\in \mathbb{X}$, and $(\hat x,p,l)\in \mathbb{\hat X}$, one gets:
	\begin{align}\notag
	\Vert \mathbb{H}(x,p,l)-\mathbb{\hat H}(\hat x,p,l)\Vert = \Vert h(x)-\hat h(\hat x)\Vert = \Vert y-\hat y\Vert.
	\end{align}
	Since $V$ is an SSF from $\mathbb{\widehat G}(\widehat \Sigma)$ to $\mathbb{G}(\Sigma)$, we have
	\begin{align}\notag
	\PP&\Big\{\sup_{0\leq k\leq T_d}\Vert y_{a\hat\nu}(k)-\hat y_{\hat a \hat\nu}(k)\Vert\geq\varepsilon\,|\,a,\hat a,p_0\Big\}\\\notag
	&=\PP\Big\{\sup_{0\leq k\leq T_d}\alpha\left(\Vert y_{a\hat \nu}(k)-\hat y_{\hat a \hat\nu}(k)\Vert\right)\geq\alpha(\varepsilon)\,|\,a,\hat a,p_0\Big\}\\\label{eq:supermart}
	&\leq\PP\Big\{\sup_{0\leq k\leq T_d} V((x_{a\hat \nu}(k),p(k),l(k),(\hat x_{\hat a\hat \nu}(k),p(k),l(k)))\geq\alpha(\varepsilon)\,|\,a,\hat a,p_0\Big\}.
	\end{align}
	The equality holds due to $\alpha$ being a $\mathcal K_\infty$ function, and also condition \eqref{eq:lowerbound2} on the SSF $V$. By applying Lemma~\ref{Lemma: Kushner} to \eqref{eq:supermart}, utilizing inequality \eqref{eq6666}, and since 
	\begin{align}\notag
	\max\Big\{\kappa V((x,p,l),(\hat x,p,l)),\psi\Big\}\leq \kappa V((x,p,l),(\hat x,p,l)) + \psi,
	\end{align}
	one can readily acquire the results in~\eqref{Eq_25}.
	
\end{IEEEproof}

\begin{IEEEproof}\textbf{(Theorem~\ref{Thm: Comp1})}
	We first show that SSF $V$ in~\eqref{Comp: Simulation Function1} satisfies the inequality \eqref{eq:lowerbound2} for some $\mathcal{K}_\infty$ function $\alpha$. For any $(x,p,l)\in \mathbb{X}$, and $(\hat x,p,l)\in \mathbb{\hat X}$, one gets:
	
	\begin{align}\notag
	\Vert \mathbb{H}(x,p,l)-\mathbb{\hat H}(\hat x,p,l)\Vert &=\max_i \{\Vert \mathbb{H}_{ii}(x_i,p_i,l_i)-\mathbb{\hat H}_{ii}(\hat x_i,p_i,l_i) \Vert\}\\\notag
	&\le\max_i \{\Vert \mathbb{H}_{i}(x_i,p_i,l_i)-\mathbb{\hat H}_{i}(\hat x_i,p_i,l_i) \Vert\}\le\max_i \{\alpha_i^{-1}(V_i((x_i,p_i,l_i),(\hat x_i,p_i,l_i)))\}\\\notag
	&\le\beta~(\max_i \{\sigma^{-1}_i(V_i((x_i,p_i,l_i),(\hat x_i,p_i,l_i)))\})=\beta(V((x,p,l),(\hat x,p,l))),
	\end{align}
	where $\beta(s)=\max_i\Big\{\alpha^{-1}_i\circ \sigma_i(s)\Big\}$ for all $s \in \mathbb R_{\ge 0}$, which is a $\mathcal{K}_\infty$ function and~\eqref{eq:lowerbound2} holds with $\alpha=\beta^{-1}$. We continue with showing that inequality~\eqref{eq6666} holds, as well.
	Let $\kappa(s)= \max_{i,j}\{\sigma_i^{-1}\circ\kappa_{ij}\circ\sigma_j(s)\}$. It follows from~\eqref{compositionality1} that $\kappa<\mathcal{I}_d$. Since $\max_{i}\sigma_i^{-1}$ is concave, one can readily get the chain of inequalities in \eqref{Equ11b} using Jensen's inequality, inequality~\eqref{eq:Pi_mu}, and by defining $\psi$ as
	\begin{align}\label{overall-error}
	\psi:=\max_{i}\sigma_i^{-1}(\Lambda_i),
	\end{align} 
	where $\Lambda_i:=(\mathcal{I}_d + \tilde \delta_f^{-1})\circ (\rho_{\mathrm{int}i}\circ \bar \lambda\circ(\bar \lambda-\mathcal{I}_d)^{-1}(\max_{j, j\neq i}\{\bar\mu_{ji}\})+\psi_i)$.
	Hence, $V$ is an SSF from $\mathbb{\widehat G}(\widehat \Sigma)$ to $\mathbb{G}(\Sigma)$, which completes the proof.
\end{IEEEproof}

\begin{remark}
	Note that to show Theorem~\ref{Thm: Comp1}, we employed the following inequalities:
	\begin{equation}\notag
	\left\{\hspace{-1.5mm}\begin{array}{l}\rho_{\mathrm{int}}(a+b)\leq\rho_{\mathrm{int}}\circ\bar \lambda(a)+\rho_{\mathrm{int}}\circ\bar \lambda\circ(\bar \lambda- \mathcal{I}_d)^{-1}(b), \\
	a+b\leq\max\{(\mathcal{I}_d+\tilde \delta_f)(a),(\mathcal{I}_d+\tilde \delta_f^{-1})(b)\},\\
	\end{array}\right.
	\end{equation}
	for any  $a,b\in\mathbb R_{\ge 0}$, where $\rho_{\mathrm{int}}, \tilde \delta_f, \bar \lambda, (\bar \lambda- \mathcal{I}_d)\in\mathcal{K}_\infty$. 
\end{remark}

\begin{remark}
	If $\rho_{\mathrm{int}i}, i\in \{1,\ldots,N\}$, are linear, $\kappa_{ij}$ and $\Lambda_i$ reduce to, respectively, $\kappa_{ij}=(\mathcal{I}_d + \tilde \delta_f)\circ\rho_{\mathrm{int}i}\circ \alpha_j^{-1}(s)$, and $\Lambda_i: =(\mathcal{I}_d + \tilde \delta_f^{-1})\circ (\rho_{\mathrm{int}i}\circ (\max_{j, j\neq i}\{\bar\mu_{ji}\})+\psi_i), \forall i\in \{1,\ldots,N\}, j\neq i$.
\end{remark}

\begin{figure*}[ht]
	\rule{\textwidth}{0.1pt}
	\begin{align}\nonumber
	\mathbb{E}&\Big[V((x',p',l'),(\hat x',p',l'))\,\big|\,x,\hat x,p,l\Big]=\mathbb{E}\Big[\max_{i}\Big\{\sigma_i^{-1}(V_i((x'_i,p'_i,l'_i),(\hat x'_i,p'_i,l'_i)))\Big\}\,\big|\,x,\hat x,p,l\Big]\\\notag
	&\le\max_{i}\Big\{\sigma_i^{-1}(\mathbb{E}\Big[V_i((x'_i,p'_i,l'_i),(\hat x'_i,p'_i,l'_i))\,\big|\, x,\hat x,p,l\Big])\Big\}=\max_{i}\Big\{\sigma_i^{-1}(\mathbb{E}\Big[V_i((x'_i,p'_i,l'_i),(\hat x'_i,p'_i,l'_i))\,\big|\,x_{i},\hat x_i,p_i,l_i\Big])\Big\}\\\notag
	&\leq\max_{i}\Big\{\sigma_i^{-1}(\max\{\kappa_iV_i((x_i,p_i,l_i),(\hat x_i,p_i,l_i)),\rho_{\mathrm{int}i}(\Vert w_{i}-\hat w_i\Vert),\psi_i\})\Big\}\\\notag
	&=\max_{i}\Big\{\sigma_i^{-1}(\max\{\kappa_iV_i((x_i,p_i,l_i),(\hat x_i,p_i,l_i)),\rho_{\mathrm{int}i}(\max_{j, j\neq i}\{\Vert w_{ij}-\hat w_{ij}\Vert\}),\psi_i\})\Big\}\\\notag
	&=\max_{i}\Big\{\sigma_i^{-1}(\max\{\kappa_iV_i((x_i,p_i,l_i),(\hat x_i,p_i,l_i)),\rho_{\mathrm{int}i}(\max_{j, j\neq i}\{\Vert y_{ji}-\hat y_{ji}+\hat y_{ji}-\Pi_{w_{ji}}(\hat y_{ji})\Vert\}),\psi_i\})\Big\}\\\notag
	&\leq\max_{i}\Big\{\sigma_i^{-1}(\max\{\kappa_iV_i((x_i,p_i,l_i),(\hat x_i,p_i,l_i)), \rho_{\mathrm{int}i}(\max_{j, j\neq i}\{\Vert \mathbb{H}_{j}\!(x_j,p_j,l_j)\!-\!\mathbb{\hat H}_{j}\!(\hat x_j,p_j,l_j)\Vert\!\!+\!\Vert\hat y_{ji}\!-\!\Pi_{w_{ji}}\!(\hat y_{ji}\!)\Vert\}\!),\psi_i\})\!\Big\}\\\notag
	&\leq\max_{i}\Big\{\sigma_i^{-1}(\max\{\kappa_iV_i((x_i,p_i,l_i),(\hat x_i,p_i,l_i)),\rho_{\mathrm{int}i}(\max_{j , j\neq i}\{\alpha_j^{-1}(V_j((x_j,p_j,l_j),(\hat x_j,p_j,l_j)))+\bar\mu_{ji}\}),\psi_i\})\Big\}\\\notag
	&\leq\max_{i}\Big\{\sigma_i^{-1}(\max\{\kappa_iV_i((x_i,p_i,l_i),(\hat x_i,p_i,l_i)),\rho_{\mathrm{int}i}\circ \bar \lambda(\max_{j, j\neq i}\{\alpha_j^{-1}(V_j((x_j,p_j,l_j),(\hat x_j,p_j,l_j)))\})\\\notag
	&~~~+\rho_{\mathrm{int}i}\circ \bar \lambda\circ(\bar \lambda-\mathcal{I}_d)^{-1}(\max_{j, j\neq i}\{\bar\mu_{ji}\}),\psi_i\})\Big\}\\\notag
	&\leq\max_{i}\Big\{\sigma_i^{-1}(\max\{\kappa_iV_i((x_i,p_i,l_i),(\hat x_i,p_i,l_i)),(\mathcal{I}_d + \tilde \delta_f)\circ\rho_{\mathrm{int}i}\circ \bar \lambda(\max_{j, j\neq i}\{\alpha_j^{-1}(V_j((x_j,p_j,l_j),(\hat x_j,p_j,l_j)\!)\!)\}),\Lambda_i\})\!\Big\}\\\notag
	&=\max_{i,j}\Big\{\sigma_i^{-1}(\max\{\kappa_{ij}(V_j((x_j,p_j,l_j),(\hat x_j,p_j,l_j)),\Lambda_i\})\Big\}\\\notag
	&=\max_{i,j}\Big\{\sigma_i^{-1}(\max\{\kappa_{ij}\!\circ\! \sigma_j\!\circ\! \sigma_j^{-1}(V_j(\!(x_j,p_j,l_j),(\hat x_j,p_j,l_j))),\Lambda_i\})\Big\}\\\notag
	&\leq\max_{i,j,{\bar j}}\Big\{\sigma_i^{-1}(\max\{\kappa_{ij}\!\circ\! \sigma_j \!\circ\! \sigma_{\bar j}^{-1}(V_{\bar j}((x_{\bar j},p_{\bar j},l_{\bar j}),(\hat x_{\bar j},p_{\bar j},l_{\bar j}))),\Lambda_i\})\Big\}\\\notag
	&=\max_{i,j}\Big\{\sigma_i^{-1}(\max\{\kappa_{ij}\!\circ\!\sigma_j(V(\!(x,p,l),(\hat x,p,l))),\Lambda_i\})\Big\}\\\label{Equ11b}
	&=\max\{\kappa V((x,p,l),(\hat x,p,l)),\psi\}.
	\end{align}
	\rule{\textwidth}{0.1pt}
\end{figure*}

\begin{IEEEproof}\textbf{(Theorem~\ref{Thm_5a})}
		Given the general assumption on $h$, since $\Sigma_p$ is \emph{incrementally input-to-state stable} ($\delta$-ISS), and from \eqref{Con55}, $\forall(x,p,l)\in \mathbb{X}$ and $\forall(\hat x,p,l)\in\mathbb{\hat X}$, we get 
	\begin{align}\notag
	\Vert \mathbb{H}(x,p,l)-\mathbb{\hat H}(\hat x,p,l) \Vert &=\Vert h(x)-\hat h(\hat x ) \Vert \leq \mathscr{L}(\Vert x-\hat x\Vert)\leq \mathscr{L}\circ \underline \alpha_p^{-1}(V_p(x,\hat x))\\\notag
	&=\mathscr{L}\circ \underline \alpha_p^{-1}({\bar\kappa_p}^{l/\epsilon}\, V((x,p,l),(\hat x,p,l))).
	\end{align}
	Since $\frac{1}{\bar\kappa_{p}^{l/\epsilon}}>1$, one can conclude that the inequality \eqref{Eq_2a} holds with $\alpha(s)=\min_p\{(\mathscr{L}\circ \underline \alpha_p^{-1}(s))^{-1}\}$, $\forall s\in \R_{\geq0}$.
	Now we show that the inequality~\eqref{Eq_3a} holds, as well. By taking the conditional expectation from \eqref{Eq65}, $\forall x\in X, \forall \hat{x} \in \hat X, \forall p \in P,\forall w \in W,\forall \hat{w} \in \hat W$, we have 
	\begin{align}\notag
	\mathbb{E}&\Big[V_p(f_p(x,w,\varsigma),\hat f_p(\hat x,\hat w,\varsigma))\big|x,\hat x, \hat \nu,w, \hat w\Big]-\mathbb{E}\Big[V_p(f_p(x,w,\varsigma),f_p(\hat{x},\hat{w},\varsigma))\big|x,\hat x,\hat \nu,w, \hat w\Big]\\\notag
	~~&\leq\mathbb{E}\Big[\gamma(\Vert\hat f_p(\hat x,\hat w,\varsigma)-f_p(\hat x,\hat w,\varsigma)\Vert)\big|x,\hat x, \hat \nu,w, \hat w\Big],
	\end{align}
	where $\hat f_p(\hat{x},\hat{w},\varsigma) = \Pi_x(f_p(\hat{x},\hat{w},\varsigma))$. Using Remark~\ref{Def154} and inequality~\eqref{eq:Pi_delta}, the above inequality reduces to
	\begin{align}\notag
	\mathbb{E}\Big[V_p(f_p(x,w,\varsigma),\hat f_p(\hat x,\hat w,\varsigma))\big|x,\hat x,  \hat \nu,w, \hat w\Big]-\mathbb{E}\Big[V_p(f_p(x,w,\varsigma),f_p(\hat{x},\hat{w},\varsigma))\big|x,\hat x, \hat \nu,w, \hat w\Big]\leq\gamma_p(\bar\delta).
	\end{align}
	Employing \eqref{Con85}, we get 
	\begin{align}\label{triangular}
	&\mathbb{E}\Big[V_p(f_p(x,w,\varsigma),\hat f_p(\hat{x},\hat{w},\varsigma))\big|x,\hat x,w, \hat w\Big]\leq\bar{\kappa}_pV_p(x,\hat x)+\bar \rho_{\mathrm{int}p}(\Vert w-\hat w\Vert)+\gamma_p(\bar\delta).
	\end{align}
	Now, in order to show that the function $V$ in~\eqref{function V} satisfies~\eqref{Eq_3a}, we should consider the different scenarios as in Definition~\ref{def: abstract gMDP}. For the first scenario ($l<k_d-1,\Vert f_p(\hat x,\hat w,\varsigma)-\hat f_p(\hat{x},\hat{w},\varsigma)\Vert \leq \bar \delta$, $p' = p$, and $l' = l+1$), using~\eqref{triangular} we have:
	\begin{align}\notag
	\mathbb{E}& \Big[V((x',p',l'),(\hat x',p',l'))\,\big|\,x,\hat{x},p,l,w,\hat w\Big]
	=\frac{1}{{\bar\kappa_{p'}}^{l'/\epsilon}}\mathbb{E} \Big[V_{p'}(x',\hat x')\big|x,\hat{x},\hat \nu,w,\hat w\Big]\\\notag
	&=\frac{1}{{\bar\kappa_{p}}^{(l+1)/\epsilon}}\mathbb{E} \Big[V_p(f_p(x,w,\varsigma),\hat f_p(\hat{x},\hat{w},\varsigma))\big|x,\hat{x},\hat \nu,w,\hat w\Big]\leq\frac{1}{{\bar\kappa_{p}}^{(l+1)/\epsilon}}\Big(\bar{\kappa}_pV_p(x,\hat x)+\bar \rho_{\mathrm{int}p}(\Vert w-\hat w\Vert)+\gamma_p(\bar\delta)\Big)\\\notag
	&=\bar \kappa_p^\frac{\epsilon-1}{\epsilon}V((x,p,l),(\hat x,p,l))+\frac{1}{\bar{\kappa}_{p}^{(l+1)/\epsilon}}\Big(\bar \rho_{\mathrm{int}p}(\Vert w-\hat w\Vert)+\gamma_p(\bar\delta)\Big)\leq \bar \kappa_p^\frac{\epsilon-1}{\epsilon}V((x,p,l),(\hat x,p,l))\\\notag
	&+\frac{1}{\bar{\kappa}_{p}^{k_d/\epsilon}}\Big(\bar \rho_{\mathrm{int}p}(\Vert w-\hat w\Vert)+\gamma_p(\bar\delta)\Big);\notag
	\end{align}
	Note that the last inequality here holds since $l<k_d-1$, and consequently, $l+1<k_d$. 
	
	For the second scenario ($l=k_d-1,\Vert f_p(\hat x,\hat w,\varsigma)-\hat f_p(\hat{x},\hat{w},\varsigma)\Vert \leq \bar\delta$, $p' = p$, and $l'=k_d-1$), we have:
	\begin{align}\notag
	\mathbb{E}& \Big[V((x',p',l'),(\hat x',p',l'))\big|x,\hat{x},p,l,w,\hat w\Big]
	=\frac{1}{{\bar\kappa_{p'}}^{l'/\epsilon}}\mathbb{E} \Big[V_{p'}(x',\hat x')\big|x,\hat{x},\hat \nu,w,\hat w\Big]\\\notag
	&=\frac{1}{{\bar\kappa_{p}}^{l/\epsilon}}\mathbb{E} \Big[V_p(f_p(x,w,\varsigma),\hat f_p(\hat{x},\hat{w},\varsigma))\big|x,\hat{x},\hat \nu,w,\hat w\Big]\leq\frac{1}{{\bar\kappa_{p}}^{l/\epsilon}}\Big(\bar{\kappa}_pV_p(x,\hat x)+\bar \rho_{\mathrm{int}p}(\Vert w-\hat w\Vert)+\gamma_p(\bar\delta)\Big)\\\notag
	&=\bar \kappa_pV((x,p,l),(\hat x,p,l))+\frac{1}{\bar{\kappa}_{p}^{k_d/\epsilon}}\Big(\bar \rho_{\mathrm{int}p}(\Vert w-\hat w\Vert)+\gamma_p(\bar\delta)\Big)\\\notag
	& \le \bar \kappa_p^\frac{\epsilon-1}{\epsilon}V((x,p,l),(\hat x,p,l))+\frac{1}{\bar{\kappa}_{p}^{k_d/\epsilon}}\Big(\bar \rho_{\mathrm{int}p}(\Vert w-\hat w\Vert)+\gamma_p(\bar\delta)\Big);
	\end{align}
	Note that the last inequality here holds since $\epsilon > 1$, and consequently, $0<\frac{\epsilon-1}{\epsilon}<1$.
	
	For the last scenario ($l=k_d-1,\Vert f_p(\hat x,\hat w,\varsigma)-\hat f_p(\hat{x},\hat{w},\varsigma)\Vert \leq \bar\delta, p'\neq p$, and $l' = 0$), using Assumption~\ref{Assume: Switching} we have:
	\begin{align}\notag
	\mathbb{E}& \Big[V((x',p',l'),(\hat x',p',l'))\big|x,\hat{x},p,l,w,\hat w\Big]
	=\frac{1}{{\bar\kappa_{p'}}^{l'/\epsilon}}\mathbb{E} \Big[V_{p'}(x',\hat x')\big|x,\hat{x},\hat \nu,w,\hat w\Big]\\\notag
	&\leq\mu\,\mathbb{E} \Big[V_p(f_p(x,w,\varsigma),\hat f_p(\hat{x},\hat{w},\varsigma))\big|x,\hat{x},\hat \nu,w,\hat w\Big]=\mu{\bar\kappa_{p}}^{(k_d-1)/\epsilon}\frac{1}{{\bar\kappa_{p}}^{l/\epsilon}}\mathbb{E} \Big[V_p(f_p(x,w,\varsigma),\hat f_p(\hat{x},\hat{w},\varsigma))\big|x,\hat{x},\hat \nu,w,\hat w\Big]\\\notag
	&\leq\mu{\bar\kappa_{p}}^{(k_d-1)/\epsilon}\frac{1}{{\bar\kappa_{p}}^{l/\epsilon}}\Big(\bar{\kappa}_pV_p(x,\hat x)+\bar \rho_{\mathrm{int}p}(\Vert w -\hat w\Vert)+\gamma_p(\bar\delta)\Big)\\\notag
	&\leq\mu\bar\kappa_{p}^{(k_d-1)/\epsilon}\bar \kappa_pV((x,p,l),(\hat x,p,l))+\mu\Big(\bar \rho_{\mathrm{int}p}(\Vert w-\hat w\Vert)+\gamma_p(\bar\delta)\Big)\\\notag
	&\le \bar \kappa_p^\frac{\epsilon-1}{\epsilon}V((x,p,l),(\hat x,p,l))+\frac{1}{\bar{\kappa}_{p}^{k_d/\epsilon}}\Big(\bar \rho_{\mathrm{int}p}(\Vert w-\hat w\Vert)+\gamma_p(\bar\delta)\Big);\notag
	\end{align}
	Note that $\forall p\in P$, $\mu\bar\kappa_{p}^{(k_d-1)/\epsilon}\leq 1$ since $\forall p\in P$, $k_d \geq \epsilon\frac{\ln(\mu)}{\ln(1/\bar\kappa_{p})}+1$. By employing a similar argument as the one in~\cite[Theorem 1]{Abdallah2017compositional}, and by defining $\bar \kappa =\max_p\{\bar \kappa_p^\frac{\epsilon-1}{\epsilon}\}$, $\bar \rho_{\mathrm{int}}(s) =\max_p\{\frac{1}{\bar{\kappa}_{p}^{k_d/\epsilon}}\bar \rho_{\mathrm{int}p}(s)\},\forall s\in\mathbb R_{\geq0},$ and $ \bar \gamma(\bar \delta) = \max_p\{\frac{1}{\bar{\kappa}_{p}^{k_d/\epsilon}} \bar\gamma_p(\bar \delta)\}$, the following inequality
	\begin{align}\notag
	\mathbb{E}& \Big[V((x',p',l'),(\hat x',p',l'))\,\big|\,x,\hat{x},p,l,w,\hat w\Big]\le\max\Big\{\tilde\kappa V((x,p,l),(\hat x,p,l)), \tilde \rho_{\mathrm{int}}(\Vert w-\hat w\Vert),\tilde \gamma\Big\}
	\end{align}
	holds for the all scenarios, where $\tilde {\kappa} =(1-(1-\tilde \pi)(1-\bar \kappa)$, $\tilde \rho_{\mathrm{int}}=(\mathcal{I}_d + \tilde {\delta}_f)\circ(\frac{1}{(1-\bar \kappa)\tilde \pi}\bar\lambda\circ\bar \rho_{\mathrm{int}})$, $\tilde \gamma=(\mathcal{I}_d + \tilde {\delta}_f^{-1})\circ(\frac{1}{(1-\bar \kappa)\tilde \pi}\circ\bar\lambda\circ(\bar\lambda - \mathcal{I}_d)^{-1}\circ \bar \gamma)$ where $\tilde {\delta}_f,\bar\lambda,$ are some arbitrarily chosen $\mathcal{K}_\infty$ functions with $\bar\lambda - \mathcal{I}_d \in \mathcal{K}_\infty$, and $0<\tilde \pi<1$, $1-\bar \kappa >0$. Hence, inequality \eqref{Eq_3a} is satisfied with $\nu=\hat{\nu}$, $\kappa=\tilde{\kappa}$, $ \rho_{int} = \tilde \rho_{int}$, and $\psi=\tilde \gamma(\bar\delta)$. Hence, $V$ is an SPSF from $\mathbb{\widehat G}(\widehat \Sigma_i)$ to $\mathbb{G}(\Sigma_i)$, which completes the proof.
\end{IEEEproof}

\begin{remark}
	If $\forall p\in P$, there exists a common $V: X\times X\to\R_{\ge0}$ satisfies  Definition~\ref{Def11} and Assumptions~\ref{Assume: Switching} and~\ref{Assume: Switching1}, then $p= p', \forall p,p' \in P$, and consequently, $V, \alpha, \bar \kappa, \bar \rho_{\mathrm{int}}$ and $\bar \gamma$ in Theorem~\ref{Thm_5a} reduce to the functions $V((x,p,l),(\hat x,p,l)) = V(x,\hat x), \alpha(s)=(\mathscr{L}_p\circ \underline \alpha_p^{-1}(s))^{-1}, \bar \rho_{\mathrm{int}}(s) = \bar \rho_{\mathrm{int}p}(s),\forall s\in\mathbb R_{\geq0}$, and constants $ \bar \kappa = \bar \kappa_p$, $\bar \gamma (\bar \delta)= \bar \gamma_p(\bar \delta)$.
\end{remark}

\begin{figure*}
	\rule{\textwidth}{0.1pt} 
	\begin{align}\notag
	&\textbf{- First Scenario}~(l<k_d-1,\Vert f(\hat x,\hat w,\varsigma)-\hat f_p(\hat{x},\hat{w},\varsigma)\Vert \leq \bar\delta, p' = p, l' = l+1)\!:\\\notag
	\mathbb{E}& \Big[V((x',p',l'),(\hat x',p',l'))\big|x,\hat{x},p,l,w,\hat w\Big]=\frac{1}{\bar\kappa_{p'}^{l'/\epsilon}}\mathbb{E} \Big[V_{p'}(x',\hat x')\big|x,\hat{x},\hat \nu,w,\hat w\Big]\\\notag 
	&=\frac{1}{\bar\kappa_{p}^{(1+l)/\epsilon}}\mathbb{E} \Big[V_p(f_p(x,w,\varsigma),\hat f_p(\hat{x},\hat{w},\varsigma))\,\big|\,x,\hat{x},\hat \nu,w,\hat w\Big]\\\notag
	&=\frac{1}{\bar\kappa_{p}^{(1+l)/\epsilon}}\Big((x-\hat x)^T\Big[(A_p+\hat\delta_p E_pF_p)^T M_p(A_p\!+\!\hat\delta_p E_pF_p)\Big](x-\hat x)+2 \Big[(x\!-\!\hat x)^T\!(A_p\!+\!\hat\delta_p E_pF_p)^T\Big] M_p\Big[\!D_p(w\!-\!\hat w)\Big]\\\notag
	&~~~~~~~~~+2 \Big[(x-\hat x)^T(A_p+\hat\delta_p E_pF_p)^T\Big]M_p\mathbb{E}\Big[\bar N_p\,\big|\,x,\hat x, \hat \nu,w, \hat w\Big]+2 \Big[(w-\hat w)^T D_p^T\Big] M_p \mathbb{E}\Big[\bar N_p\,\big|\,x,\hat x, \hat \nu,w, \hat w\Big]\\\notag
	&~~~~~~~~~+(w-\hat w)^T D_p^T M_pD_p(w-\hat w)+\mathbb{E}\Big[\bar N_p^T M_p \bar N_p\,\big|\,x,\hat x, \hat \nu,w, \hat w\Big]\Big)\\\notag
	&\le\frac{1}{\bar\kappa_{p}^{(1+l)/\epsilon}}\Big(\begin{bmatrix}x-\hat x\\\hat\delta_p F_p(x-\hat x)\\\end{bmatrix}^T\begin{bmatrix}
	(1+2\pi_p)A_p^T M_pA_p& A_p^T M_pE_p\\
	*& (1+2\pi_p)E_p^T  M_pE_p
	\end{bmatrix}\begin{bmatrix}x-\hat x\\\hat\delta_p F_p(x-\hat x)\\\end{bmatrix}\\\notag
	&~~~~~~~~~+\bar p(1+\pi_p+2/\pi_p){\Vert\sqrt{M_p}D_p\Vert_2^2}\Vert w-\hat w\Vert^2+n(1+3/\pi_p)\lambda_{\max}{(M_p)}\,\bar\delta^2\Big)\\\notag
	&\le\frac{1}{\bar\kappa_{p}^{(1+l)/\epsilon}}\Big(\begin{bmatrix}x-\hat x\\\hat\delta_p F_p(x-\hat x)\\\end{bmatrix}^T\begin{bmatrix}
	\bar\kappa_p M_p& -F_p^T\\
	-F_p & \frac{2}{\bar a_p}
	\end{bmatrix}\begin{bmatrix}x-\hat x\\\hat\delta_p F_p(x-\hat x)\\\end{bmatrix}+\bar p(1+\pi_p+2/\pi_p){\Vert\sqrt{M_p}D_p\Vert_2^2}\Vert w-\hat w\Vert^2 \\\notag
	&~~~~~~~~~+n(1+3/\pi_p)\lambda_{\max}{(M_p)}\,\bar\delta^2\Big)\\\notag
	&=\frac{1}{\bar\kappa_{p}^{(1+l)/\epsilon}}\Big(\bar\kappa_p (V_p(x,\hat x))-2\hat\delta_p(1-\frac{\bar\delta}{\bar a_p})(x-\hat x)^TF_p^T F_p( x-\hat x)+\bar p(1+\pi_p+2/\pi_p){\Vert\sqrt{M_p}D_p\Vert_2^2}\Vert w-\hat w\Vert^2 \\\notag
	&~~~~~~~~~+ n(1+3/\pi_p)\lambda_{\max}{(M_p)}\,\bar\delta^2\Big)\\\notag
	&\le\bar \kappa_p^\frac{\epsilon-1}{\epsilon} V((x,p,l),(\hat x,p,l))+\frac{1}{\bar{\kappa}_{p}^{k_d/\epsilon}}\Big(\bar p(1+\pi_p+2/\pi_p)\Vert\sqrt{M_p}D_p\Vert_2^2\Vert w-\hat w\Vert^2 +n(1+3/\pi_p)\lambda_{\max}{(M_p)}\,\bar\delta^2\Big);\\\notag
	&\textbf{- Second Scenario}~(l=k_d-1,\Vert f(\hat x,\hat w,\varsigma)-\hat f_p(\hat{x},\hat{w},\varsigma)\Vert \leq \bar\delta, p' = p, l'=k_d-1)\!:\\\notag
	\mathbb{E}& \Big[V((x',p',l'),(\hat x',p',l'))\,\big|\,x,\hat{x},p,l,w,\hat w\Big]=\frac{1}{\bar\kappa_{p'}^{l'/\epsilon}}\mathbb{E} \Big[V_{p'}(x',\hat x')\,\big|\,x,\hat{x},\hat \nu,w,\hat w\Big]\\\notag
	&=\frac{1}{\bar\kappa_{p}^{l/\epsilon}}\mathbb{E} \Big[V_p(f_p(x,w,\varsigma),\hat f_p(\hat{x},\hat{w},\varsigma))\,\big|\,x,\hat{x},\hat \nu,w,\hat w\Big]\\\notag
	&\le\bar \kappa_pV((x,p,l),(\hat x,p,l))+\frac{1}{\bar{\kappa}_{p}^{k_d/\epsilon}}\Big(\bar p(1+\pi_p+2/\pi_p)\Vert\sqrt{M_p}D_p\Vert_2^2\Vert w-\hat w\Vert^2 +n(1+3/\pi_p)\lambda_{\max}{(M_p)}\,\bar\delta^2\Big)\\\notag
	&\le\bar \kappa_p^\frac{\epsilon-1}{\epsilon} V((x,p,l),(\hat x,p,l))+\frac{1}{\bar{\kappa}_{p}^{k_d/\epsilon}}\Big(\bar p(1+\pi_p+2/\pi_p)\Vert\sqrt{M_p}D_p\Vert_2^2\Vert w-\hat w\Vert^2 +n(1+3/\pi_p)\lambda_{\max}{(M_p)}\,\bar\delta^2\Big);\\\notag
	&\textbf{- Last Scenario}~(l=k_d-1,\Vert f(\hat x,\hat w,\varsigma)-\hat f_p(\hat{x},\hat{w},\varsigma)\Vert \leq \bar\delta, p'\neq p, l' = 0)\!:\\\notag
	\mathbb{E}& \Big[V((x',p',l'),(\hat x',p',l'))\,\big|\,x,\hat{x},p,l,w,\hat w\Big]=\frac{1}{\bar\kappa_{p'}^{l'/\epsilon}}\mathbb{E} \Big[V_{p'}(x',\hat x')\,\big|\,x,\hat{x},\hat \nu,w,\hat w\Big]\\\notag
	&=\mu~\mathbb{E} \Big[V_p(f_p(x,w,\varsigma),\hat f_p(\hat{x},\hat{w},\varsigma))\,\big|\,x,\hat{x},\hat \nu,w,\hat w\Big]\\\notag
	&\le\mu\bar\kappa_{p}^{(k_d-1)/\epsilon}\bar \kappa_pV((x,p,l),(\hat x,p,l))+\mu\Big(\bar p(1+\pi_p+2/\pi_p)\Vert\sqrt{M_p}D_p\Vert_2^2\Vert w-\hat w\Vert^2 +n(1+3/\pi_p)\lambda_{\max}{(M_p)}\,\bar\delta^2\Big)\\\label{Eq_305a}
	&\le\bar \kappa_p^\frac{\epsilon-1}{\epsilon} V((x,p,l),(\hat x,p,l))+\frac{1}{\bar{\kappa}_{p}^{k_d/\epsilon}}\Big(\bar p(1+\pi_p+2/\pi_p)\Vert\sqrt{M_p}D_p\Vert_2^2\Vert w-\hat w\Vert^2 +n(1+3/\pi_p)\lambda_{\max}{(M_p)}\,\bar\delta^2\Big).
	\end{align}
	\rule{\textwidth}{0.1pt}
\end{figure*}

\begin{IEEEproof}\textbf{(Theorem~\ref{Thm_3a})}
	Since $\hat C=C$, we have $\Vert \mathbb{H}(x,p,l)-\mathbb{\hat H}(\hat x,p,l)\Vert = \Vert C x-\hat C\hat x\Vert^2\leq n\lambda_{\max}(C^TC)\Vert x\\- \hat x\Vert^2$, and similarly  $\lambda_{\min}(M_p)\Vert x- \hat x\Vert^2\leq(x-\hat x)^T M_p(x-\hat x)$. One can readily verify that  $\frac{\lambda_{\min}(M_p)}{n\lambda_{\max}(C^TC)}\Vert C x-\hat C\hat x\Vert^2\le V_p(x,\hat x)$ holds $\forall x$, $\forall \hat x$, and consequently, $\frac{1}{\bar\kappa_{p}^{l/\epsilon}}\frac{\lambda_{\min}(M_p)}{n\lambda_{\max}(C^TC)}\Vert C x-\hat C\hat x\Vert^2\le V((x,p,l),(\hat x,p,l))$, $\forall(x,p,l)\in \mathbb{X}, \forall(\hat x,p,l)\in \mathbb{\hat X}$. Since $\frac{1}{\bar\kappa_{p}^{l/\epsilon}}>1$, one can conclude that inequality~\eqref{Eq_2a} holds with $\alpha(s)=\min_p\{\frac{\lambda_{\min}(M_p)}{n\lambda_{\max}(C^TC)}\}\,s^2$ for any $s\in\mathbb R_{\geq0}$. We proceed with showing that the inequality~\eqref{Eq_3a} holds, as well. We simplify
\begin{align}\notag
A_px &+E_p\varphi_p(F_px)+ B_p+ D_pw +R_p\varsigma-\Pi_x(A_p\hat x + E_p\varphi_p(F_p\hat x)+B_p + D_p\hat w + R_p\varsigma)
\end{align}
to 
\begin{align}\label{Eq_552}
A_p(x-\hat x)+ D_p(w-\hat w) +E_p(\varphi_p(F_px)-\varphi_p(F_p\hat x))+\bar N_p,
\end{align}
where $\bar N_p =  A_p\hat x +E_p\varphi_p(F_p\hat x)+B_p+ D_p\hat w + R_p\varsigma-\Pi_x(A_p\hat x+E_p\varphi_p(F_p\hat x)+ B_p+ D_p\hat w + R_p\varsigma)$.
From the slope restriction~\eqref{Eq_6a}, one obtains
\begin{align}\label{Eq_129a}
\varphi_p(F_px)-\varphi_p(F_p\hat x)=\hat\delta_p(F_px-F_p\hat x)=\hat\delta_p F_p(x-\hat x),
\end{align}
where $\hat\delta_p$ is a function of $x$ and $\hat x$ and takes values in the interval $[0,\bar a_p]$. Using~\eqref{Eq_129a}, the expression in~\eqref{Eq_552} reduces to
\begin{align}\notag
(A_p+\hat \delta_p E_pF_p)(x-\hat x)+D_p(w-\hat w)+\bar N_p.
\end{align}
Using Young's inequality~\cite{young1912classes} as $cd\leq \frac{\pi}{2}c^2+\frac{1}{2\pi}d^2,$ for any $c,d\geq0$ and any $\pi>0$, by employing Cauchy-Schwarz inequality and~\eqref{Eq_88a}, and since 
\begin{align}\notag
\Vert \bar N_p\Vert~\leq~ \bar\delta,\quad \bar N_p^T M_p \bar N_p \leq n\lambda_{\max}(M_p)\bar\delta^2,
\end{align}
one can obtain the chain of inequalities in~\eqref{Eq_305a} including the different scenarios as in Definition~\ref{def: abstract gMDP}. By employing the similar argument as the one in~\cite[Theorem 1]{Abdallah2017compositional}, and by defining $\bar \kappa =\max_p\{\bar \kappa_p^\frac{\epsilon-1}{\epsilon}\}$, $\bar \rho_{\mathrm{int}}(s) =\max_p\{\frac{1}{\bar{\kappa}_{p}^{k_d/\epsilon}}\bar p(1+\pi_p+2/\pi_p)\Vert\sqrt{M_p}D_p\Vert_2^2\}s^2,\forall s\in\mathbb R_{\geq0},$ and $ \bar \gamma = \max_p\{\frac{1}{\bar{\kappa}_{p}^{k_d/\epsilon}} n(1+3/\pi_p)\lambda_{\max}{(M_p)}\}\bar \delta^2$, the following inequality
\begin{align}\label{max form}
\mathbb{E} \Big[V((x',p',l'),(\hat x',p',l'))\,\big|\,x,\hat{x},p,l,w,\hat w\Big]\le\max\Big\{\tilde\kappa V((x,p,l),(\hat x,p,l)), \tilde \rho_{\mathrm{int}}(\Vert w-\hat w\Vert),\tilde \gamma\Big\}
\end{align}
holds for all the scenarios, where $\tilde {\kappa} =(1-(1-\tilde \pi)(1-\bar \kappa)$, $\tilde \rho_{\mathrm{int}}=\frac{(1+\tilde \delta_c) }{(1-\bar \kappa) \tilde\pi}\bar \rho_{\mathrm{int}}$, $\tilde \gamma=\frac{(1+1/\tilde \delta_c) }{(1-\bar \kappa) \tilde\pi}\bar \gamma$, where $\tilde \pi,\tilde \delta_c,$ can be arbitrarily chosen such that $0<\tilde \pi<1$, $\tilde \delta_c >0$, $1-\bar \kappa >0$. Therefore, the inequality \eqref{Eq_3a} is satisfied with $\nu=\hat{\nu}$, $\kappa=\tilde{\kappa}$, $ \rho_{int} = \tilde \rho_{int}$, and $\psi=\tilde \gamma$. Hence, $V$ defined in~\eqref{function V: Nonlinear} is an SPSF from $\mathbb{\widehat G}(\widehat \Sigma)$ to $\mathbb{G}(\Sigma)$, which completes the proof.
\end{IEEEproof}

\end{document}